\documentclass[a4paper,12pt]{article}
\usepackage{amsmath,amssymb}
\usepackage[top=3cm,bottom=3cm,left=3cm,right=3cm]{geometry}
\usepackage{color}
\usepackage[T1]{fontenc}
\usepackage{hyperref}
\hypersetup{colorlinks=true}
\usepackage{cite}
\date{}
\begin{document}
\author{Haci  Akbas ${^\dag}$ and O. Teoman Turgut$^\ddag$ \\ Department of Physics, Bo\u{g}azi\c{c}i University \\ 34342 Bebek, Istanbul, Turkey \\ $^\dag$akbas@gmail.com, $^\ddag$turgutte@boun.edu.tr}
\title{\bf Born-Oppenheimer Approximation for a Singular System}
\maketitle
\begin{abstract}
We discuss a  simple singular system in one dimension, two heavy particles interacting with a light particle via an attractive contact interaction. It is natural to apply Born-Oppenheimer approximation to this problem. We present a detailed discussion of this approach, the advantage of this simple model is that one can estimate the error terms self-consistently. Moreover, a Fock space approach to this problem is presented where a systematic expansion can be proposed to get higher order corrections. A slight modification of the same problem in which the light particle is relativistic is discussed in a later section. Here, the second quantized description is more challenging but with some care one can recover the first order term as well as introducing a more systematic approach to higher orders.
\end{abstract}
\section{Introduction}
Born-Oppenheimer approximation \cite{Born-Opp} is the basic tool  of molecular physics \cite{landau, bethe-jackiw,weinbergQM}, it is also fundamental in solids, for example in  modelling coupling of lattice vibrations to electronic degrees of freedom (for a rather detailed coverage of various aspects of this field, one may look for example at \cite{mahan}, there are also classics on the subject such as \cite{born-huang, ziman}). In the case of molecular physics, where all particles interact via  Coulomb forces, the Born-Oppenheimer vibrational energy levels go with $(m/M)^{1/2}$, where $m/M$ refers to the  light mass to heavy mass ratio, rotational energy levels as well as anharmonic corrections go with $(m/M)$, therefore they are  at a higher order, the relevant expansion parameter being considered as $(m/M)^{1/4}$. Our main concern here  is essentially the static Born-Oppenheimer approximation in which one is interested in stationary levels of the system. There is a large literature on this subject, we will not be able to cover all of it, we only mention some  works aiming at a rigorous approach to the Born-Oppenheimer approximation, which is somewhat related to our desire to get some control over the error terms in our toy model. We are not discussing the time-dependent Born-Oppenheimer approximation, this is a very interesting and closely related subject,  the reader can consult the review articles \cite{Spohn, Hagedorn-Joye, Jecko-5} for more information.

An interesting toy model worked out by R. Seiler in \cite{seiler-2}, where two heavy and one light particle all interact via harmonic oscillator potentials. In this work it is verified that the assumptions of the Born-Oppenheimer approach hold. Following this, some  rigorous aspects of Born-Oppeheimer approximation is presented in \cite{combes-seiler}. It is not at all clear that the eigenvalues of the light degrees of freedom, that one computes, under the influence of potentials when the heavy centers are clamped,  actually define well-behaved nonintersecting surfaces  when one considers the heavy degrees of freedoms as parameters. This difficult problem is solved by Hunziker in \cite{hunziker}, where even for Coulomb type potentials energy eigenfunctions are shown to be essentially analytic functions of the heavy coordinates. These problems further investigated in a series of papers by Hagedorn \cite{hagedorn4,hagedorn5,hagedorn6}. An attempt to include higher order corrections to Born-Oppenheimer approximation is given by Weingert and Littlejohn \cite{weigert} as an example of their diagonalization technique in the deformation quantization approach. They discover derivative terms in the corrections and it is consistent  with what we find here as well. Higher order corrections also were rigorously investigated by Hagedorn in a series of papers \cite{hagedorn1,hagedorn2}, they are essentially angular momentum and nonlinear oscillation terms, as predicted originally by Born-Oppenheimer. Further investigations along similar lines are presented in \cite{seiler-klein}. The reader can find a large collection references and mathematically precise statements on Born-Oppenheimer approximation in a recent review by Jecko \cite{Jecko-5}

In a similar spirit to \cite{seiler-2}, a slightly simpler model for its pedagocial value was proposed by G. Gangopadhyay and B. Dutta-Roy in \cite{dutta-roy} where the authors consider a light particle coupled to a heavy particle via a delta function potential, which makes it slightly singular, and the whole system is confined to a box in one dimension for which analytical treatment is possible. Our toy model investigates Born-Openheimer approximation in a very similar singular system, albeit leading to  finite results again thanks to being formulated in one dimension, but it is physically more interesting. The simplicity of the model allows us to test various aspects and higher order corrections to Born-Oppenheimer approximation. We consider two heavy particles interacting with a light particle through an attractive delta function potential. As a result of the attraction the heavy particles form a kind of molecule, but a singular one, since there is no repulsion they collapse onto one another if we consider them as classical particles on the ground state energy surface.  When the two heavy particles are separated by a small distance there is an effective linear attractive potential acting on them. This leads to a linear oscillator that one can solve exactly.
In our problem, we note that the relevant expansion parameter  is $(m/M)^{1/3}$, different from the usual molecular systems. The consistency of our approximations are verified in the first Appendix, by computing the order of each neglected term, using  the proposed solution for the estimates. Moreover, we note that the higher order corrections can be introduced in a second quantized language, using an approach suggested by Rajeev\cite{rajeev-asymptotic}, which contain some higher order derivative terms, not so simple to identify as in the case of rotational degrees of freedom of molecules. These are some of the novel aspects of this problem. The relativistic dispersion relation for the light particle could be important to gain  some insight into the heavy quark systems, in which the gluons, being massles always to be treated relativistically leading to a  linear attractive potential between the heavy quarks. Here we study the one dimensional version of the model studied previsously for two dimensions in \cite{caglar-teo} to understand a proper finite formulation via a nonperturbative renormalization process. Again we find a linear effective potential as a result of the interactions with the  light particle, the derivation of which is given in Appendix-II.  What would be more interesting is to study the two dimensional version of the present problem where there are renormalization issues to be taken care of before one employs the Born-Oppenheimer type approximations\cite{akbas-teo}.

\section{Born-Oppenheimer for  Delta Function Potentials}

In this section we apply  the conventional Born-Oppenheimer approximation to a very simple model in one-dimension. Let us consider two heavy particles, each one of  which is  interacting with a light particle through an attractive delta function potential. 
The Hamiltonian of the system can be written as, 
\begin{equation}
\Big[-\frac{\hbar^2}{2M}\sum_{i}{\nabla_i}^2-\frac{\hbar^2}{2m}\nabla^2-\lambda\delta(x-x_1)-\lambda\delta(x-x_2)]\Big]\Psi(x;x_1, x_2)=E\Psi(x;x_1,x_2).
\end{equation}
Here $x_1,x_2$ refer to the heavy particles coordinates and $x$ refers to the light one. The choices of the masses also reflect this difference. 
Let us assume that the Born-Oppenheimer approximation can be applied to this system, that is we introduce  a decomposition of the wave function into fast and
slow degrees of freedom:
\begin{equation}
\Psi(x;x_1, x_2)=\phi(x|x_1, x_2)\psi(x_1,x_2).
\end{equation}
We assume that this decomposition respects the translational invariance of the system and we will make use of this when we estimate the error terms. We can always assume that the wave function has such a decomposition. Note that the conventional assumption would be that the wave function can be expanded into a  series of  complete eigenfunctions of the light-degree of freedom with coefficients depending on the heavy degrees of freedom.
After completing our calculations, we will see that the derivative of the wave function $ \phi $ with respect to the large mass coordinates $ x_i $ will give us smaller terms, which will entail us to decouple the light degree of freedom from the heavy ones. We substitute the proposed solution into  the Schr\"odinger equation, 
\begin{eqnarray}
&\ & \Big[-\frac{\hbar^2}{2M}\sum_{i}{\nabla_i}^2\psi(x_1,x_2)\Big]\phi(x|x_1,x_2)+ \Big[-\frac{\hbar^2}{2M}\sum_{i}{\nabla_i}^2\phi(x|x_1,x_2)\Big]\psi(x_1,x_2) \nonumber\\
&\ & -\frac{\hbar^2}{2M}\sum_{i} {\partial \phi\over \partial x_i}{\partial \psi\over \partial x_i}
  +\Big[\big(-\frac{\hbar^2}{2m}\nabla_x^2-\lambda\delta(x-x_1)-\lambda\delta(x-x_2)\big)\phi(x |x_1,x_2)\Big]\psi(x_1, x_2)\nonumber\\
&\  & \quad \quad \quad \quad\ \ \ \ \ \ \ \ \ \ \ \ \ \ \ \ \ \ \  \ \ \ \ \ \ \ \ \ \ \ \quad =E\phi(x|x_1,x_2)\psi(x_1,x_2).
\end{eqnarray}
Let us assume that we find the solution to the equation below,
\begin{equation}
-\frac{\hbar^2}{2m}\nabla_x^2 \phi(x|x_1, x_2)-\lambda[\delta(x-x_1)+\delta(x-x_2)]\phi(x|x_1, x_2) =E(x_1,x_2)\phi(x|x_1, x_2)
.\end{equation}
This has a simple interpretation, we assume that the heavy particles act like fixed centers and the light particle moves in this background.
As a result we find,
\begin{eqnarray}
&\ & \Big[-\frac{\hbar^2}{2M}\sum_{i}{\nabla_i}^2+E(x_1,x_2)\Big]\psi(x_1,x_2)\phi(x|x_1,x_2)+ \Big[-\frac{\hbar^2}{2M}\sum_{i}{\nabla_i}^2\phi(x|x_1,x_2)\Big]\psi(x_1,x_2) \nonumber\\
&\ & -\frac{\hbar^2}{2M}\sum_{i} {\partial \phi\over \partial x_i}{\partial \psi\over \partial x_i} =E\phi(x|x_1,x_2)\psi(x_1,x_2).
\end{eqnarray}
Therefore, if we can neglect the last two terms on the lefthand side, we end up with the Born-Oppenheimer result,
\begin{equation}
\Big[-\frac{\hbar^2}{2M}\sum_{i}{\nabla_i}^2+E(x_1,x_2)\Big]\psi(x_1,x_2)=E\psi(x_1,x_2)
.\end{equation} 
Thus the solution of  equation (4) for the light degrees of freedom generates  an effective potential for the heavy degrees of freedom, which we can then solve in principle. 
Let us recall the solution to this  fixed center problem,  the heavy degrees of freedom are frozen at their locations, thus  a particle is moving under the influence of two delta potentials. 
 This is a standart quantum mechanics problem, which we can easily solve by the following approach, inspired from the two dimensional version of delta function potentials \cite{fatih-teo}.
Let us make the following ansatz for the ground state wave-function 
\begin{equation}
\phi(x|x_1, x_2) = \sum_{i=1}^2\int_0^\infty {dt\over \hbar}  K_t(x,x_i)e^{-\frac{\nu^2}{\hbar}t},
\end{equation}
here $K_t(x,y)$ refers to the heat kernel on $\bf{R}$, which is simply a Gaussian. The heat kernel satisfies the well-known heat equation,
\begin{equation}
-\frac{\hbar^2}{2m}\nabla^2K_t(x,y) +\hbar\frac{\partial K_t(x,y)}{\partial t}= 0
,\end{equation}
the solution of which is unique under the conditions,
\begin{equation}
K_t(x,y)=K_t(y,x),\quad{\rm and} \qquad \lim_{t\to 0^+} K_t(x,y)=\delta(x-y)
,\end{equation}
moreover we impose  $K_t(x,y)\geq 0$, which is  required for the solutions in more general contexts. 
One can see that the solution in flat space  is given by a Gaussian,
\begin{equation}
K_t(x,y) =\sqrt{\frac{2m}{4\pi\hbar t}}e^{-\frac{2m}{4\hbar t}|x-y|^2}.
\end{equation} 
The proposed wave function is  positive and symmetric as required by the properties of the ground state wave function and the symmetry of the problem due to equal strength delta functions. The antisymmetric combination is of a higher energy level.
We verify that this ansatz  indeed solves the equation with a negative eigenvalue,
if and only if $E(x_1,x_2)=-\nu^2$ satisfies
\begin{equation}
\frac{1}{\lambda} -\frac{1}{\hbar}\int_0^\infty dt K_t(x_2,x_2)e^{-\frac{\nu^2}{\hbar} t}=\frac{1}{\hbar}\int_0^\infty{dt}K_t(x_1,x_2)e^{-\frac{\nu^2 t}{\hbar}}
\end{equation}
 this claim can be verified by   the following integration by parts trick, 
\begin{eqnarray}
     \int_0^\infty {dt\over \hbar} \Big(-{\hbar^2\over 2m}\Big){\partial^2\over \partial x^2} K_t(x,x_i)e^{-\frac{\nu^2}{\hbar} t}& =&-\int_0^\infty {dt\over \hbar} \hbar{\partial\over \partial t} K_t(x,x_i)e^{-\frac{\nu^2}{\hbar} t}\nonumber\\
& =&-\int_0^\infty dt \Big({\partial\over \partial t} (K_t(x,x_i)e^{-\frac{\nu^2}{\hbar} t})-K_t(x,x_i)e^{-\frac{\nu^2}{\hbar} t}\Big)\nonumber\\
&=& \delta(x-x_i)-(-\nu^2) \int_0^\infty {dt\over \hbar} K_t(x,x_i)e^{-\frac{\nu^2}{\hbar} t} \nonumber
,\end{eqnarray}
where we used the inital condition on the heat kernel and its boundedness as $t\to \infty$.
 One can see that for any given positive value of $\lambda$ there is always a solution for $\nu$. 
 Moreover, one can  verify  that the above  solution agrees with the usual solution one would find by the Fourier transform method. Nevertheless, this representation of the solution is more useful for our calculations, and it can be generalized to the case of curves embedded in a Riemannian manifold.
When we place the two delta functions onto the same location,   that is $z=|x_1-x_2|=0$, we find
\begin{equation}
\frac{1}{2\lambda} = \frac{1}{\hbar}\int_0^\infty dt K_t(x_1,x_1)e^{-\frac{\nu_0^2}{\hbar}t}=\sqrt{\frac{m}{2\hbar^2\nu_0^2}}.
\end{equation}
As a result we find for the zeroth order energy, 
\begin{equation}
\nu_0^2= \frac{{2m}\lambda^2}{\hbar^2}.
\end{equation}
Normally, we expect a nonzero distance between the two centers, in this case the soltion is found from the full expression. One would see immediately that the energy  achieves minimum when the distance between the centers is zero, that corresponds to the  equlibrium configuration for the heavy system. When we displace them slightly from this equlibrium configuration, we may calculate the resulting energy change, which would act like an effective potential for the slow degrees of freedom in the Born-Oppenheimer approach. So under the assumption that small values of $z$ make the main contribution to the dynamics, we get a small correction to the energy, $E(x_1,x_2)=-\nu_0^2 +\Delta E(z)$, when we insert this back again into the equations,
\begin{eqnarray}
\frac{1}{\lambda} -\sqrt{\frac{m}{2\pi\hbar}}\int_0^\infty \frac{dt }{\hbar}\frac{e^{-\frac{\nu^2-\Delta E}{\hbar}t}}{\sqrt{t}}&=&\sqrt{\frac{m}{2\pi\hbar}}\int_0^\infty\frac{dt}{\hbar}\frac{e^{-\frac{(\nu^2-\Delta E_1)}{\hbar}t-\frac{mz^2}{2\hbar t}}}{\sqrt{t}}\nonumber\\
&=& \sqrt{\frac{m}{2\hbar^2(\nu_0^2-\Delta E)}}e^{\frac{|z|}{\hbar}\sqrt{2m(\nu_0^2-\Delta E)}}
\end{eqnarray}
We could  neglect $\Delta E|z|/\nu_0 $ terms since both $\Delta E$ and $z$  are assumed small,  as we will verify, we then find,
\begin{equation}
\frac1{\lambda}-\sqrt{\frac{m}{2\hbar^2(\nu_0^2-\Delta E)}} =\sqrt{\frac{m}{2\hbar^2(\nu_0^2-\Delta E)}}e^{-\frac{\nu_0 |z|\sqrt{2m}}{\hbar}}
\end{equation}
and expanding everything in the same order leads to the following  relation between 
$\Delta E $ and $z$: 
\begin{equation}
\Delta E= \lambda^3|z|\Big(\frac{ 2m}{\hbar^2}\Big)^2
.\end{equation}
We are now ready for the Born-Oppenheimer approach, we introduce this energy as an effective potential acting between the heavy particles, since it depends on the separation between them. This gives us the following Schr\"odinger equation:
\begin{equation}
-\frac{\hbar^2}{2\mu}{\partial^2\over \partial z^2}\psi +|z|\lambda^3\left(\frac{2m}{\hbar^2}\right)^2\psi =\delta E\psi,
\end{equation}
here we represent the excitation energies, corresponding to $E+\nu_0^2$, by $\delta E$, also we use the reduced mass  $\mu=M_1M_2/(M_1+M_2)=M/2$ since only the relative coordinate appears in the equation.
Note that this is a particle under the influence of a linear potential, the solutions of which are well-known, an especially beautiful presentation can be found in \cite{schwinger}. 
Let us define the variables below,
\begin{eqnarray}
\beta^3 &=& (2\mu)\left(\frac{2m}{\hbar^3}\right)^2\lambda^3,\quad {\rm or}\quad \beta=\Big({\mu\over m}\Big)^{1/3}{\nu_0^2\over \lambda} \quad{\rm also,}\nonumber\\
u &=&|z| -\left(\frac{\hbar^2}{2m}\right)^2\frac{\delta E}{\lambda^3}\ \  {\rm and}\ \ 
\sigma = \beta u
.\end{eqnarray}
Note that $\lambda$ has dimensions energy-length.
As a result we find the solutions given by the Airy functions, being separated into even or odd ones,
\begin{equation}
\psi_{\pm}(z) =C\Big(\begin{matrix}{\rm sgn}(z)\\ 1\end{matrix}\Big) Ai\left((2\mu)^\frac1{3}\frac{(2m)^\frac{2}{3}}{\hbar^2}\lambda\left(|z| - \left(\frac{\hbar^2}{2m}\right)^2\frac{\delta E}{\lambda^3}\right)\right)
.\end{equation}
If we now impose the continuity of the wave functions and their derivates at
 $z$= 0, we find, for the odd and even respectively, 
either the zeros of $Ai(\sigma)$ or $Ai'(\sigma)$, which we collectively denote by $-\sigma_n$ with $n$ odd refering to the odd and $n$ even refering to the even solutions \cite{schwinger}. This leads to the following quantization conditions for the eigenvalues of  the linear oscillator,
\begin{equation}
\delta E_n=(-\sigma_n) {2 m\lambda^2\over \hbar^2}\Big({m\over \mu}\Big)^{1/3}=(-\sigma_n)\nu_0^2\Big({m\over \mu}\Big)^{1/3},
\end{equation}
which shows explicitly that $\delta E_n << \nu_0^2$.
We can now accomplish the normalization of each wave function, 
which can be turned into the following integral,
\begin{equation}
2{ C_n}^2\int_{\sigma_n}^\infty d\sigma| Ai(\sigma)|^2 = 1
.\end{equation}
This integral can be evaluated explicitly, giving us,
\begin{equation}
\frac{1}{2{ C_n}^2} = \left(\frac{\partial Ai(\sigma)}{\partial\sigma}\right)^2|_{\sigma=\sigma_n}-\sigma_n Ai(\sigma_n)^2
.\end{equation}
As a result, for $n$ is  even, we get
\begin{equation}
\frac{1}{2{ C_n}^2} = -\sigma_n Ai(\sigma_n)^2
\end{equation}
and similarly, for $n$ is  odd, we have,
\begin{equation}
\frac{1}{2{ C_n}^2}= \left(\frac{\partial Ai(\sigma)}{\partial\sigma}\right)^2|_{\sigma=\sigma_n}
.\end{equation}
As a result we find the wave function,
\begin{equation}
\psi_n(z) = C_n({\mu\over m})^{1/6} {\nu_0\over \lambda^{1/2}}\Big(\begin{matrix}{\rm sgn}(z)\\ 1\end{matrix}\Big) Ai\left((2\mu)^\frac1{3}\frac{(2m)^\frac{2}{3}}{\hbar^2}\lambda\left(|z| - \left(\frac{\hbar^2}{2m}\right)^2\frac{\delta E_n}{\lambda^3}\right)\right)
,\end{equation}
indeed  here $({\mu\over m})^{1/6} {\nu_0\over \lambda^{1/2}}=(2\mu)^{1/6}{(2 m)^{1/3}\lambda^{1/2}\over \hbar}$.
If we assume that we have two identical bosonic heavy particles, we only need to consider the even wave functions which are symmetric under the interchange of these two particles, which corresponds to $z\mapsto-z$ here. 
This essentially completes the discussion on Born-Oppenheimer approximation, apart from checking the consistency of our approximations. For this,  we will calculate the expectation value of the variable $|z|$, it must be of the same order as the kinetic energy, and we find that 
\begin{eqnarray}
(2{C_n}^2)\int_{\sigma_n}^\infty d\sigma \left( \frac{\sigma\hbar^2}{(2\mu)^\frac1{3}(2m)^\frac{2}{3}\lambda} + \left(\frac{\hbar^2}{2m}\right)^2\frac{\delta E}{\lambda^3}\right)|Ai(\sigma)|^2 &=& -\frac{2}{3}\frac{{\sigma_n}}{\lambda }\frac{\hbar^2}{(2\mu)^\frac{1}{3}(2m)^\frac{2}{3}}\nonumber\\
&=&\Big[-\frac{2}{3}\sigma_n\Big] \frac{\lambda}{\nu_0^2}\Big(\frac{m}{\mu}\Big)^{1/3}.\nonumber
\end{eqnarray}
Next, to ensure that  higher order terms in the potential are negligible,  we will calculate the spread of the wave function, by evaluating the expectation value of $z^2$. 
This can be done,  based on  the fomulae given in \cite{Airy}, we calculate  $< z^2>$  as follows,
\begin{equation}
< z^2> = 2{C_n}^2\int_{\sigma_n}^\infty d\sigma |Ai(\sigma)|^2\left( \frac{\sigma\hbar^2}{(2\mu)^\frac1{3}(2m)^\frac{2}{3}\lambda} + \left(\frac{\hbar^2}{2m}\right)^2\frac{\delta E}{\lambda^3}\right)^2
\end{equation}
we can calculate this integral term by term, as an example we have,
\begin{equation}
 \frac{\hbar^4}{(2\mu)^\frac{2}{3}(2m)^\frac{4}{3}\lambda^2}(2{C_n}^2)\int_{\sigma_n}^\infty d\sigma {\sigma}^2|Ai(\sigma)|^2  =\frac1{5}\frac{{\sigma_n}^2}{\lambda^2}\frac{\hbar^4}{(2\mu)^\frac{2}{3}(2m)^\frac{4}{3}}\left(1-\frac1{{\sigma_n}^3}\right)
,\end{equation}
and we only quote the result for $n$ even, since we are mainly interested in the identical particle case,
which leads to the following result, 
$<z^2>$ ($n$ even case) : 
\begin{equation}
<z^2> =\frac{8}{15}\frac{{\sigma_n}^2}{\lambda^2}\frac{\hbar^4}{(2\mu)^\frac{2}{3}(2m)^\frac{4}{3}} -\frac1{5}\frac{\hbar^4}{(2\mu)^\frac{2}{3}(2m)^\frac{4}{3}\lambda^2 \sigma_n}=A_n \Big({m\over \mu}\Big)^{2/3}\Big( {\lambda\over \nu_0^2}\Big)^2
,\end{equation}
 where $A_n$ is a numerical factor.

Having found this solution, we may go back and check the consistency of these approximations. This is presented in the Appendix-I, where we show for the proposed solution, that  indeed the terms we neglect, lead to smaller order corrections.

\section{A Many Body View}
We will approach the same problem from the many-body perspective, construct the principal operator for a set of interacting particles, the Hamiltonian of which is given as follows:
\begin{equation}
H=\int dx \phi^\dagger (x) [-{\nabla^2\over 2m}] \phi(x)+\int dx \psi^\dagger(x)  [-{\nabla^2\over 2M}] \psi(x)-\lambda \int dx \phi^\dagger (x) \psi^\dagger(x) \phi(x)\psi(x)
\end{equation}
 For the extension,  we will use the algebra of orthofermions as suggested by Rajeev in\cite{rajeev-asymptotic} (named as angels in Rajeev's work). 
We therefore modify the problem using an extended Fock space construction. This is accomplished by an algebra defined through a set of operation rules, 
\begin{eqnarray} \chi(x)\chi^{\dagger}(y) &=&  \delta(x-y) \Pi_0,\cr\nonumber
\chi_(x)\chi(y)=0&=&
\chi^{\dag}(x)\chi^{\dag} (y) ,
\end{eqnarray}
where
\begin{equation} \Pi_1 = \int dx  \,
\chi^{\dag}(x) \chi(x) , \;\;\;\Pi_0 = 1- \Pi_1 \; 
\end{equation}
are the projection operators onto the one-angel and no-angel states,
respectively. This algebra has a realization on ${\bf C}\oplus {\cal L}^2({\bf R})$.
Thus we extend the Hilbert space of the theory  from ${\cal  F}_M\otimes {\cal F}_m$ to 
${\cal  F}_M\otimes {\cal F}_m\oplus {\cal  F}_M\otimes {\cal F}_m\otimes L^2({\bf R})$.

Define a new Hamiltonian on this extended space  in matrix form  as follows
\begin{eqnarray}
&\ &\hat H-E\Pi_0=\left (
\begin{array}{cc}
(H_0-E) \Pi_0 &  \int dx  \psi^\dag(x)\phi^\dag (x) \chi(x) \\
\int dy \psi(y) \phi(y) \chi^\dag(y) &  {1\over \lambda}\Pi_1 \\
\end{array} \right )\equiv \left (
\begin{array}{cc}
a & b^\dagger \\
b & d \\
\end{array} \right )
\end{eqnarray}
The resolvent or the Green's function of this extended system is defined as
\begin{equation}
(\hat H-E \Pi_0)^{-1} \equiv \left (
\begin{array}{cc}
\alpha & \beta^\dagger \\
\beta & \delta \\
\end{array} \right )
\end{equation}
The projection of this matrix Green's function on to the no-angel (no-orthofermion) subspace can be fromally written in two alternative ways:
\begin{eqnarray}
\nonumber \alpha = (a-b^\dagger d^{-1} b)^{-1} = (H-E)^{-1} 
=a^{-1}+a^{-1}b^\dagger \Phi^{-1}ba^{-1},
\end{eqnarray}
where we  introduce the principal matrix $\Phi$, given by
\begin{equation}
\Phi \equiv d-ba^{-1}b^\dagger
.\end{equation}
The second expression becomes more useful in our calculations.
The first  relation for $\alpha$, by the properties of the orthofermion operators,  shows that the projection of the resolvent of the new operator onto the no-angel subspace reproduces the Green's function of the original Hamiltonian. In the same way therefore the second representation also reproduces the Green's function.
In our case we have the explicit expression,
\begin{eqnarray}
\nonumber \Phi = {1\over \lambda}  \Pi_1 -\int dx \phi(x)\psi(x)\chi^\dag(x) {1\over H_0-E}\int dy \psi^\dag (y) \phi^\dag (y) \chi(y)\nonumber
\end{eqnarray}
Let us normal order this operator, the result, can be written in the momentum representation. We warn the reader that {\it we use the following notational convention}, $[dp]={dp\over 2\pi}$ and 
delta function, written in momentum space as $\delta[p-q]$ refers to $2\pi\delta(p-q)$, having established that, the normal ordered operator becomes,
\begin{eqnarray}
\Phi&=&{1\over \lambda} \Pi_1-\int [dpdq] \chi^\dagger(p+q){1\over H_0+\nu_0^2-\delta' E+p^2/2M+q^2/2m}\chi(p+q)\nonumber\\
&-&\int [dpdqdr] \chi^\dag(p+q)\psi^\dag(r){1\over H_0+\nu_0^2
-\delta' E+p^2/2M+r^2/2M+q^2/2m}\psi(p) \chi(r+q),\nonumber
\end{eqnarray}
where we dropped a term which contains normally ordered light particles since we are assuming that there is only a single light particle. Moreover, we use again the splitting of the energy as $-\nu_0^2+\delta' E$.  In this formalism two heavy and a single light is replaced with an orthofermion and a single heavy particle. 

In the first term change to the center of momentum and relative momentum,
$P=p+q$ and $\eta=\mu[p/M-q/m]$, where $\mu$ is the reduced mass, the first term becomes:
\begin{equation}
\int [dPd\eta] \chi^\dagger(P){1\over H_0+\nu_0^2-\delta' E+P^2/2(M+m)+\eta^2/2\mu}\chi(P)
\end{equation}
Integral over the relative momenta can be executed;
\begin{equation} 
{1\over 2\hbar}\sqrt{{2mM\over m+M}}\int [dP] \chi^\dagger(P){1\over \sqrt{H_0+\nu_0^2-\delta' E+P^2/2(M+m)}}\chi(P)
.\end{equation}
The key idea, due to Rajeev is  this, the bound state solutions can only come from the zero eigenvalues of the $\Phi(E)$ operator.
If we are interested in the bound states of this system we will look for an eigenfunction, $|\omega>$,
\begin{equation}
\Phi(E) |\omega>=0
,\end{equation}
this normalizable solution $|\omega>$ can be used to obtain the actual bound state solution easily.
In $H_0$ we have both the light and heavy particle free Hamiltonians, but recall that we have no light particle when we switch to the principal operator in this sector, that means $H_0=H_0(\psi)$. We  thus  assume that $H_0-\delta' E$ in the kinetic term, and $p^2/2M+r^2/2M-\delta' E$ in the potential term, are small relative to 
$\nu_0^2$ type  terms. Everything can be expanded relative to the large  $\nu_0^2$ term or $\nu_0^2+q^2/2m$ term respectively, depending on the kinetic or potential terms, and we drop the extra term ${m/M}$ in the kinetic energy as well as in the reduced mass. The reason behind this complication can be understood as follows, we cannot assume that $\delta' E$ is small, because there is also the total kinetic energy of center of mass motion, which could be large, but it is also included in the other kinetic energy terms, as we will see in a moment, hence only their combined sum can be small. It turns out that  to get everything consistent we need to do a second order expansion, as we will see.
\begin{eqnarray}
\Phi&=& \frac1{\lambda}\Pi_1-\frac{\sqrt{2m}}{2\hbar\nu_0}\int [dP]\chi^\dagger(P)\left[1-\frac1{2}\frac{(H_0-\delta' E+\frac{P^2}{2M})}{\nu_0^2}+\frac{3}{8}\frac{(H_0-\delta' E+\frac{P^2}{2M})^2}{\nu_0^4}+...\right]\chi(P)\nonumber\\
&-&\!\!\!\!\int [dpdqdr]\chi^\dagger(p+q)\psi^\dagger(r)\left[\frac1{(\nu_0^2+\frac{q^2}{2m})}-\frac{(\frac{p^2}{2M}+\frac{r^2}{2M}-\delta' E)}{(\nu_0^2+\frac{q^2}{2m})^2}+\frac{(\frac{p^2}{2M}+\frac{r^2}{2M}-\delta' E)^2}{(\nu_0^2+\frac{q^2}{2m})^3}+...\right]\nonumber\\
&\ & \qquad \qquad \qquad \qquad \qquad \qquad \times \psi(p)\chi(q+r)\nonumber
\end{eqnarray}
Using this expansion for $\Phi$,  we make the following ansatz for the zero eigenvalue solution of $\Phi$ operator. 
\begin{equation}
|\omega> =\int [d\xi dQ]f(\xi)e^{-iQX/\hbar}\chi^\dagger(\frac{Q}{2}+\xi)\psi^\dagger(\frac{Q}{2}-\xi)|\Omega>,
\end{equation}
where, $|\Omega>$ denotes the vacuum for the combined Fock space of particles and orthofermion.  Here we may assume that the relative wave function in position space is real, this implies a symmetry for the Fourier transform, $f(\xi)=f^*(-\xi)$.  Note that in principle we can improve this ansatz by taking into account the fact that the orthofermion is a composite of light and heavy, hence its mass is actually $M+m$, but this is a smaller order improvement, hence we may ignore it.  This approximation actually implies an approximate symmetry for the wave function, up to order $m/M$, we may assume that due to the bosonic nature of the heavy particles,  the wave function $f(z)$ is ivariant under inversion, i. e. $f(z)=f(-z)$ (this symmetry should be violated by terms of order $m/M$). This means that  Fourier transform satisfies $f(\xi)=f(-\xi)$. 

  Let us now compute the action of $\Phi$ on our ansatz. The first part of $\Phi$ operator produces,
\begin{eqnarray}
&\ &\Big(\frac1{\lambda}-\frac1{2}\frac{\sqrt{2m}}{\hbar\nu_0}\Big)\Pi_1|\omega>+\frac1{4}\frac{\sqrt{2m}}{\hbar\nu^3}\int [dP] \chi^\dagger(P)\big(H_0-\delta' E+\frac{P^2}{2M}\big)\chi(P)|\omega>\nonumber\\
&\ &=\Big(\frac1{\lambda}-\frac1{2}\frac{\sqrt{2m}}{\hbar\nu_0}\Big)\Pi_1|\omega>+\frac{1}{4}\frac{\sqrt{2m}}{\hbar\nu^3}\int [dQd\xi] f(\xi)e^{-i{QX\over \hbar}}\big(\frac1{2}\frac{Q^2}{2M}+2\frac{\xi^2}{2M}-\delta' E\big)\nonumber\\
&\ & \qquad\qquad\qquad \qquad\qquad\qquad \qquad \times \chi^\dagger(\frac{Q}{2}+\xi)\psi^\dagger(\frac{Q}{2}-\xi)|\Omega>
,\end{eqnarray}
using the convolution to express the particle-orthofermion Fock state in relative coordinate space and then stripping off the Fock state vector, we have the expression,
\begin{eqnarray}
\left(\frac1{\lambda}-\frac1{2}\frac{\sqrt{2m}}{\hbar\nu_0}\right)f(z) +\frac1{4}\frac{\sqrt{2m}}{\hbar\nu^3}\left[\frac1{2}\frac{Q^2}{2M}-\frac{\hbar^2}{2\mu}\nabla^2_z-\delta' E\right]f(z)
.\end{eqnarray}
Similarly we work on the "potential" part of the $\Phi$ operator, the first term of which is given by
\begin{eqnarray}
&\ &\int [dpdqdr]\chi^\dagger(p+q)\psi^\dagger(r)\frac1{\nu_0^2+\frac{q^2}{2m}}\psi(p)\chi(q+r)|\omega>\nonumber\\
&\ & =\int[ dqdQd\xi] f(\xi)e^{-iQX/\hbar}\frac1{\nu_0^2+\frac{q^2}{2m}}\chi^\dagger(\frac{Q}{2}-\xi +q)\psi^\dagger(\frac{Q}{2}+\xi -q)|\Omega>.
\end{eqnarray}
Here, we redefine $\xi \rightarrow -\xi $ and $ \xi \rightarrow \xi -q $
\begin{equation}
\int [dqdQd\xi] f(q-\xi)e^{-iQX/\hbar}\frac1{\nu_0^2+\frac{q^2}{2m}}\chi^\dagger(\frac{Q}{2}+\xi )\psi^\dagger(\frac{Q}{2}-\xi )|\Omega>
\end{equation}
and assuming the ground state wave function $f(\xi)$  is symetric so  $ f(q-\xi) \rightarrow f(\xi-q)$ 
the final result can be turned into an expression for $f(z)$ only;
\begin{equation}
\int [dqdQd\xi] f(q-\xi)e^{-iQX/\hbar}\frac1{\nu_0^2+\frac{q^2}{2m}}\chi^\dagger(\frac{Q}{2}+\xi )\psi^\dagger(\frac{Q}{2}-\xi )|\Omega> \mapsto {1\over 2}\frac{\sqrt{2m}}{\hbar\nu_0}e^{-\frac{\sqrt{2m}}{\hbar}\nu_0 |z|}f(z)
.\end{equation}
Thus, combining the first result with this one, we have,
\begin{equation}
\left(\frac1{\lambda}-\frac1{2}\frac{\sqrt{2m}}{\hbar\nu_0}\right)f(z)+\frac1{4}\frac{\sqrt{2m}}{\hbar\nu_0^3}\left[\frac1{2}\frac{Q^2}{2M}-\frac{\hbar^2}{2\mu}\nabla^2_z -\delta' E\right]f(z)-\frac1{2}\frac{\sqrt{2m}}{\hbar\nu_0}e^{-\frac{\sqrt{2m}}{\hbar}\nu_0 |z|}f(z)
.\end{equation}
There is one more term coming from the potential part, this last term acting on $|\omega>$ gives us,
\begin{eqnarray}
&\ &\!\!\!\!\!\!\!\!\!\!\!\int[ dpdqdr]\chi^\dagger(p+q)\psi^\dagger(r)\frac{(\frac{p^2}{2M}+\frac{r^2}{2M}-\delta' E)}{(\nu_0^2+\frac{q^2}{2m})^2}\psi(p)\chi(q+r)|\omega>\nonumber\\
&\ &=\int [dqdQd\xi] f(\xi)e^{-iQX/\hbar}\left(\frac{\frac{(\frac{Q}{2}-\xi)^2}{2M}+\frac{(\frac{Q}{2}+\xi-q)^2}{2M}-\delta' E}{(\nu_0^2+\frac{q^2}{2m})^2}\right)\chi^\dagger(\frac{Q}{2}-\xi+q)\psi^\dagger(\frac{Q}{2}+\xi-q)|\Omega>\nonumber\\
&\ &=\int [dqdQd\xi ]f(\xi-q)e^{-iQX/\hbar}\left(\frac{\frac{(\frac{Q}{2}+\xi-q)^2}{2M}+\frac{(\frac{Q}{2}-\xi)^2}{2M}-\delta' E}{(\nu_0^2+\frac{q^2}{2m})^2}\right)\chi^\dagger(\frac{Q}{2}+\xi)\psi^\dagger(\frac{Q}{2}-\xi)|\Omega>
.\end{eqnarray}
Note that if we consider the expression below,
\begin{equation}
\frac{({Q\over 2} +\xi-q)^2}{2M}=\frac{({Q\over 2}+\xi)^2}{2M}\underbrace{-\frac{2m}{2M}\frac{2q({Q\over 2}+\xi)}{2m}+\left(\frac{2m}{2M}\right)\frac{q^2}{2m}}_{(*)}
,\end{equation}
inside the above kernel, the second and the last terms, denoted by $(*)$ collectively here,  are negligible corrections, if we ignore these two terms, we  restore the symmetric structure of this kernel.
Let us remark that this is the same order of magnitude approximation as  assuming the bosonic inversion symmetry with respect to the  relative coordinate of the orthofermion and the heavy particle in our ansatz and they are clearly related, the equations essentially signal that the assumed symmetry cannot be exact. Let us briefly digress on  this issue,
an exact calculation of this term leads to the following expression,
\begin{eqnarray}
(*)&\mapsto& -{2m\over M}\left(\frac1{4}\frac{\left(\frac{Q}{2}-i\hbar\frac{\partial}{\partial z}\right)}{2m}\frac{\hbar}{i}\frac{\partial}{\partial z}+\frac1{8}\frac{\hbar^2}{2m}\frac{\partial^2}{\partial z^2}\right)\frac{\sqrt{2m}}{\hbar\nu_0^3}e^{-\frac{\sqrt{2m}}{\hbar}\nu_0 |z|}\left[1+\frac{\sqrt{2m}}{\hbar}\nu_0 |z|\right]f(z)\nonumber
\end{eqnarray}
which explicitly shows that it is of order $m/M$, and it contains terms which will break the reflection symmetry of $f(z)$ via mixing with the center of mass momentum $Q$. Note that, $Q$ is left undetermined, assuming the system moves  with a well-defined value, but in reality this should also be smeared out with some function $g(Q)$, and every $Q$-term leads to $-i\hbar {\partial \over \partial X}$ acting on the wave function $g(X)$ in the coupled set of equations. A more complete treatment should take this into account. An estimate of this term can be made by using the first  order solution we have in the previous section and it produces,
\begin{eqnarray}
{\nu_0^2\over 8}\Big(\frac{\sqrt{2m}}{\hbar\nu_0^3}\Big)\left[\frac{iQ}{\sqrt{2m}\nu_0}\Bigg(O\left(\frac{m}{\mu}\right)^\frac{4}{3}+O\left(\frac{m}{\mu}\right)^\frac{5}{3}+...\Bigg)+\Bigg(O\left(\frac{m}{\mu}\right)+O\left(\frac{m}{\mu}\right)^\frac{4}{3}+O\left(\frac{m}{\mu}\right)^\frac{5}{3}+....\Bigg)\right]\nonumber
.\end{eqnarray}
Here, we deliberately kept the coefficients in front of it  in an unsimplified form, since in further calculations, these coefficients show up in all the terms. Note that $Q$ term above  actually brings a contribution   at a higher order. In principle, we ignore this term and then check the consistency after we find the first order solution, which will agree with the above estimate.
Hence, after ignoring these terms,  the final result of this part becomes,
\begin{eqnarray}
&\ & \int [dqdQd\xi] f(\xi-q)e^{-iQX/\hbar}\left(\frac{\frac{(\frac{Q}{2}+\xi)^2}{2M}+\frac{(\frac{Q}{2}-\xi)^2}{2M}-\delta' E}{(\nu_0^2+\frac{q^2}{2m})^2}\right)\chi^\dagger(\frac{Q}{2}+\xi)\psi^\dagger(\frac{Q}{2}-\xi)|\Omega>\mapsto\nonumber\\
&\ & \qquad \qquad \qquad \mapsto\frac1{4}\frac{\sqrt{2m}}{\hbar\nu_0^3}\left[\frac1{2}\frac{Q^2}{2M}-\frac{\hbar^2}{2\mu}\nabla^2_z -\delta' E \right]e^{-\frac{\sqrt{2m}}{\hbar}\nu_0 |z|}\Big[1+\frac{\sqrt{2m}}{\hbar}\nu_0 |z|\Big]f(z)
\nonumber
\end{eqnarray}
We remark that {\it there is another valid choice for this term}, that will be obtained by adding an extra $q$ into it, giving us,
\begin{eqnarray}
&\ &\int [dqdQd\xi] f(\xi)e^{-iQX/\hbar}\left(\frac{\frac{(\frac{Q}{2}-\xi)^2}{2M}+\frac{(\frac{Q}{2}+\xi-q)^2}{2M}-\delta' E}{(\nu_0^2+\frac{q^2}{2m})^2}\right)\chi^\dagger(\frac{Q}{2}-\xi+q)\psi^\dagger(\frac{Q}{2}+\xi-q)|\Omega>\nonumber\\
&\ &=\int [dqdQd\xi ]f(\xi-q)e^{-iQX/\hbar}\left(\frac{\frac{(\frac{Q}{2}+\xi-q)^2}{2M}+\frac{(\frac{Q}{2}-\xi+q)^2}{2M}-\delta' E}{(\nu_0^2+\frac{q^2}{2m})^2}\right)\chi^\dagger(\frac{Q}{2}+\xi)\psi^\dagger(\frac{Q}{2}-\xi)|\Omega>\nonumber
.\end{eqnarray} 
The correct expression is then found by taking the half sum of these terms, since that means the  integral expression is actually a symmetric kernel, that is the usual ordering prescription for noncommuting operators, essentially leading to expressions like,
\begin{equation}
(-\frac{\hbar^2}{2\mu}\nabla^2_z) V(z)+V(z)(-\frac{\hbar^2}{2\mu}\nabla^2_z )
.\end{equation}
The corrections coming from this non-commuting expressions are of order $m/M$, as we will verify shortly.

As a result we combine all the first order in the  kinetic energy terms, which implies that  we find the  general solution for the eigenvector $|\omega>$, to this order (yet we keep the symmetric structure),  from the equation,
\begin{eqnarray}
&\ &\left(\frac1{\lambda}-\frac1{2}\frac{\sqrt{2m}}{\hbar\nu_0}\right)f(z)+\frac1{4}\frac{\sqrt{2m}}{\hbar\nu_0^3}\left[\frac1{2}\frac{Q^2}{2M}-\frac{\hbar^2}{2\mu}\nabla^2_z -\delta' E\right]f(z)-\frac1{2}\frac{\sqrt{2m}}{\hbar\nu_0}e^{-\frac{\sqrt{2m}}{\hbar}\nu_0 |z|}f(z)\nonumber\\
&\ &\qquad\qquad +\frac{1}{8}\frac{\sqrt{2m}}{\hbar\nu_0^3}\left[\Big[\frac{1}{2}\frac{Q^2}{2M}-\frac{\hbar^2}{2\mu}\nabla^2_z -\delta' E\Big],e^{-\frac{\sqrt{2m}}{\hbar}\nu_0 |z|}\Big[1+\frac{\sqrt{2m}}{\hbar}\nu_0 |z|\Big]\right]_+f(z)=0\nonumber
.\end{eqnarray}
Note that the anti-commutator is not important to this order, it is kept to emphasize the hermitian nature of the problem.
Now, {\it we assume that higher orders in $z$ are also small}, the consistency of which can be verified after the solution, so  we expand this equation in $z$ to find the first order solution. 
So the final equation one considers for this system becomes
\begin{equation}
\left(\frac1{\lambda}-\frac{\sqrt{2m}}{\hbar\nu_0} \right)f(z) +\frac1{2}\frac{\sqrt{2m}}{\hbar\nu_0^3}\left[\frac1{2}\frac{Q^2}{2M}-\frac{\hbar^2}{2\mu}\nabla^2_z -\delta' E \right]f(z) +\frac1{2}\frac{2m}{\hbar^2}|z|f(z) =0 
\end{equation}
If we solve this system order by order, the zeroth order term gives us 
\begin{equation}
\nu_0^2= {2m\lambda^2\over \hbar^2}
,\end{equation}
which then can be inserted into the remaining part to give us the equation;
$$
\left(-\frac{\hbar^2}{2\mu}\nabla^2_z -\Big[\delta' E-\frac1{2}\frac{Q^2}{2M}\Big] \right)f(z) + \frac{\sqrt{2m}}{\hbar}\nu_0^3 |z| f(z) = 0, 
$$
or after introducing the small energy difference $\delta E=\delta' E-{1\over 2} {Q^2\over 2M}$, can be  equivalently rewritten as,
$$
\left(-\frac{\hbar^2}{2\mu}\nabla^2_z -\delta E \right)f(z) + \left(\frac{2m}{\hbar^2}\right)^2\lambda^3 |z| f(z) = 0,
$$
which is exactly the equation found by the Born-Oppenheimer approximation in the previous section.

Having found the solution, an immediate computation of the actual symmetrized term reveals that, the ordering ambiguity brings the following contribution, 
\begin{eqnarray}
&\ &\!\!\!\!\!\!\!\!\!\!\!\frac1{8}\frac{\sqrt{2m}}{\hbar\nu_0^3}(2C_n^2)\int_{\sigma_n}^\infty d\sigma Ai(\sigma)\left(-\frac{\hbar^2}{2\mu}\nabla^2_z e^{-\frac{\sqrt{2m}}{\hbar}\nu_0| z|}\Big[\frac{\sqrt{2m}}{\hbar}\nu_0 |z|+1\Big]\right)Ai(\sigma)\nonumber\\
&= & -\frac{\sqrt{2m}}{8\hbar\nu_0^3}\frac{\hbar^2}{2\mu}(2C_n^2)\int_{\sigma_n}^\infty d\sigma Ai(\sigma)\left(\nabla^2_z\Big[-\frac1{2}\frac{2m}{\hbar^2}\nu_0^2 z^2+\frac1{3}\frac{(2m)^\frac{3}{2}}{\hbar^3}\nu_0^3|z|^3-\frac1{8}\frac{(2m)^2}{\hbar^4}\nu_0^4z^4+...\Big]\right) Ai(\sigma)\nonumber\\
&=&-\frac{\sqrt{2m}}{8\hbar\nu_0^3}\frac{\hbar^2}{2\mu}(2C_n^2)\int_{\sigma_n}^\infty d\sigma Ai(\sigma)\left(-\frac{2m}{\hbar^2}\nu_0^2+2\frac{(2m)^\frac{3}{2}}{\hbar^3}\nu_0^3|z|-\frac{3}{2}\frac{(2m)^2}{\hbar^4}\nu_0^4z^2\right)Ai(\sigma)\nonumber\\
& =&{\sqrt{2m}\over \hbar\nu_0^3}\left[O\left(\nu_0^2\left(\frac{m}{\mu}\right)\right)+O\left(\nu_0^2\left(\frac{m}{\mu}\right)^\frac{4}{3}\right)+O\left(\nu_0^2\left(\frac{m}{\mu}\right)^\frac{5}{3}\right)+...\right]\nonumber
,\end{eqnarray}
where $\sigma$ is the shifted coordinate and $\sigma_n<0$ is the root of the Airy function corresponding to the energy level, $C_n$ is the normalization constant as defined in the previous section. Hence, we verify that  this ordering ambiguity has no effect at this order as claimed.

This raises the possibility of obtaining higher order corrections, which we know to be of order $({m\over M})^{2/3}$ from our previous estimates. Note that to find these terms, we may still neglect terms of order ${m\over M}$, so the reflection symmetry of $f(z)$ can be kept. To find the next order correction we resort to perturbation theory. We indicate the general approach and leave some of the details to the reader.
We assume that $\delta E$ can be expanded as $\delta_1 E+\delta_2 E+...$ where the terms represent contributions of order $({m\over M})^{1/3}, ({m\over M})^{2/3}...$ respectively. Moreover $\Phi$ itself has such an expansion too, that we should  consider.
Let us then expand  the operator $\Phi$  further,
\begin{eqnarray}
\Phi&=&\frac1{\lambda}\Pi_1 -\frac{\sqrt{2m}}{\hbar\nu_0}\int [dP]\chi^\dagger(P)\left[1-\frac1{2}\frac{(H_0-\delta_1 E -\delta_2 E+\frac{P^2}{2M})}{\nu_0^2}+\frac{3}{8}\frac{(H_0-\delta_1 E+\frac{P^2}{2M})^2}{\nu_0^4}+...\right]\chi(P)
\nonumber\\
&-&\int [dpdqdr]\chi^\dagger(p+q)\psi^\dagger(r)\left[\frac1{\nu_0^2+\frac{q^2}{2m}}-\frac{(\frac{p^2}{2M}+\frac{r^2}{2M}-\delta_1 E -\delta_2 E)}{(\nu_0^2+\frac{q^2}{2m})^2}+\frac{(\frac{p^2}{2M}+\frac{r^2}{2M}-\delta_1 E)^2}{(\nu_0^2+\frac{q^2}{2m})^3}+..\right]\nonumber \\
&\ & \qquad \qquad\qquad \qquad \qquad \qquad \qquad \qquad \qquad \qquad \qquad \qquad \qquad \times \psi(p)\chi(q+r)\nonumber
\end{eqnarray}
The first order term $\delta_1 E$ is found exactly as above, the eigenvector to this order is called $|\omega_1>$. Note that the zeroth order does not determine the eigenvector, it is a scalar identity. The next order term can be found by using {\it  first order perturbation theory}  on the eigenvalue equation. This comes from our defining equation $\Phi(E)|\omega>=\omega(E)|\omega>$. Here the equation $\omega(\nu_0+\delta_1E+\delta_2E)+\delta_2\omega(\nu_0+\delta_1 E)=0$ should be solved.
The variation of the eigenvalue can be found from first order perturbation theory, the change of the first order part  is found by the Feynmann-Helmann theorem.
Note that, we should now {\it expand the potential terms to higher orders to improve our solution to first order equation, so that it  includes the  next order corrections as well}. This implies $f(z)$, hence its eigenvalue,  should further be corrected in the first order solution.
Thus, we should compute,
\begin{eqnarray}
<\omega_1|\delta_2 \Phi|\omega_1>\!\!&+&\!\!<\omega_1|{\partial\Phi\over \partial E}|\omega_1>\delta_2E =-\frac1{4}\frac{\sqrt{2m}}{\hbar\nu_0^3}\delta_2 E-\frac1{4}\frac{(2m)^\frac{3}{2}}{\hbar^3}\nu_0<z^2>\nonumber\\
&\ & -\frac{3}{16}\frac{\sqrt{2m}}{\hbar\nu_0^5}\int [dP]<\omega_1|\chi^\dagger(P)
\left(H_0-\delta_1 E +\frac{P^2}{2M}\right)^2\chi(P)|\omega_1>\nonumber\\
&\ &-\int [dpdqdr] <\omega_1|\chi^\dagger(p+q)\psi^\dagger(r)\frac{(\delta_2 E)}{(\nu_0^2+\frac{q^2}{2m})^2}\psi(p)\chi(q+r)|\omega_1>\nonumber\\
& \ &-\int [dpdqdr] <\omega_1|\chi^\dagger(p+q)\psi^\dagger(r)\frac{(\frac{p^2}{2M}+\frac{r^2}{2M}-\delta_1 E)^2}{(\nu_0^2+\frac{q^2}{2m})^3}\psi(p)\chi(q+r)|\omega_1>\nonumber
.\end{eqnarray}
Note that in the third and the  last terms we did not keep $\delta_2 E$. Setting the whole expression equal to zero will determine the unknown $\delta_2 E$. Here $<z^2>$ again refers to the expectation with respect to the Airy functions we have found in the previous section, as a result,
the second term can be easily computed.
The kinetic energy correction can be found as,
\begin{eqnarray}
(*)&=&-\frac{3}{16}\frac{\sqrt{2m}}{\hbar\nu_0^5}\int [dP]<\omega_1|\chi^\dagger(P)\left(H_0-\delta_1 E +\frac{P^2}{2M}\right)^2\chi(P)|\omega_1>\nonumber\\
 & =& -\frac{3}{16}\frac{\sqrt{2m}}{\hbar\nu_0^5}(2 C^2_n)\int_{\sigma_n}^\infty d\sigma Ai(\sigma)\left(\frac{\hbar^4}{(2\mu)^2}\nabla^4_z+2\delta_1 E\frac{\hbar^2}{2\mu}\nabla^2_z+\delta_1 E^2\right)Ai(\sigma)\nonumber
,\end{eqnarray}
where $\delta_1 E$ could be at its $n$th level, as we discussed previously, $\sigma_n<0$ refers to the corresponding root of the Airy function again.
Let us recall that $\sigma$ was defined through,
\begin{equation}
\frac{\sigma\hbar^2}{(2m)^\frac{2}{3}(2\mu)^\frac1{3}\lambda}+\frac{\delta_1 E}{\lambda^3}\left(\frac{\hbar^2}{2m}\right)^2 =z 
,\end{equation}
and this root in turn determines the energy level, 
\begin{equation}
-\sigma_n =(2\mu)^\frac1{3}\frac{\hbar^2}{(2m)^\frac{4}{3}\lambda^2}\delta_1 E^{(n)}
.\end{equation}
We suppress this dependence of $\delta_1 E$  on $n$ for simplicity. This computation can be done and the result comes out to be,
\begin{equation}
(*)=-a_n\left(\frac{m}{\mu}\right)^\frac{2}{3}\frac1{\lambda} =-a_n\frac{\sqrt{2m}}{\hbar\nu_0}\left(\frac{m}{\mu}\right)^\frac{2}{3}
\end{equation}
where $a_n$ is a known numerical factor. 
The expectation value of $z^2$ has  already been computed,
\begin{equation}
-\frac{1}{4}\frac{(2m)^\frac{3}{2}}{\hbar^3}\nu_0<z^2> =-b_n\left(\frac{m}{\mu}\right)^\frac{2}{3}\frac1{\lambda}=-b_n\frac{\sqrt{2m}}{\hbar\nu_0}\left(\frac{m}{\mu}\right)^\frac{2}{3}
.\end{equation}
The next  term has already been computed, we only need to add $\delta_2 E$ to it, 
\begin{equation}
(**)=-\int[ dpdqdr] <\omega_1|\chi^\dagger(p+q)\psi^\dagger(r)\frac{\delta_2 E}{(\nu_0^2+\frac{q^2}{2m})^2}\psi(p)\chi(q+r)|\omega_1>
\end{equation}
which leads to,
\begin{equation}
(**)=-\frac{1}{4}\frac{\sqrt{2m}}{\hbar\nu_0^3}\delta_2 E
\end{equation}
We may now compute the last term, 
\begin{equation}
(***)=-\int [dpdqdr]<\omega_1|\chi^\dagger(p+q)\psi^\dagger(r)\frac{\left(\frac{p^2}{2M}+\frac{r^2}{2M}-\delta_1 E\right)^2}{(\nu_0^2+\frac{q^2}{2m})^3}\psi(p)\chi(q+r)|\omega_1>
.\end{equation}
To this order of accuracy, this term can also be symmetrized, to be fully consistent one may choose a Weyl ordering, that will lead to terms of the form,
\begin{equation}
({-\hbar^2\over 2\mu} \nabla^2_z )^2 V(z)+{-\hbar^2\over 2\mu} \nabla^2_z V(z){-\hbar^2\over 2\mu} \nabla^2_z +V(z)({-\hbar^2\over 2\mu} \nabla^2_z )^2
,\end{equation}
by doing a calculation similar to our consistency check, one can see that the ordering ambiguities will bring terms of order $m/M$ hence can be neglected at this order. Derivatives acting on the Airy functions will give us contributions at the desired order, they can all be computed.
As a result, one finds,
$$
(***)=-d_n\frac{\sqrt{2m}}{\hbar\nu_0}\left(\frac{m}{\mu}\right)^\frac{2}{3},
$$
where $d_n$ is an explicitly computable constant. Adding all the contributions and then  setting them to zero, we see that 
one can find the second order energy expression,
$$
\delta_2 E^{(n)}=-\alpha_n\nu_0^2\left(\frac{m}{\mu}\right)^\frac{2}{3}
$$
where $\alpha_n$ is a known purely numerical constant.
This completes our discussion on the nonrelativistic case.

\section{Relativistic Treatment of the  Light Particle}

Here we will introduce a slightly modified version of the previous problem, where the light particle is treated with a relativistic dispersion relation. This is in a sense similar to the heavy quark problem, where gauge fields binding the heavy quarks  should be considered relativistic, yet the resulting effective heavy quark system is nonrelativistic. 
First let us  define the Hamiltonian of the system, where the light particle binding is so strong that its binding energy  is comparable to its mass, hence  we should treat it relativistically. Yet the coupling is not so strong to cause pair creation. On the other hand the heavy particles are so much heavier compared to the light one that the effect of the point  interactions on them can be treated nonrelativistically. This can be modeled by a non-local many-body Hamiltonian,
\begin{equation} 
H={1\over 2}\int dx :\pi^2+\phi (-\nabla^2+m^2)\phi :+\int dx \Psi^\dagger(x) \Big( -{\nabla^2\over 2M}\Big)\Psi(x)-\lambda \int \phi^{(-)}(x)\Psi^\dagger(x)\Psi(x) \phi^{(+)}(x) dx\nonumber
\end{equation}
We follow the usual convention of setting $\hbar=1$ and $c=1$ in this section.
The light particle can be assumed to have  no charge so it is simply represented by a real field. Quantization of this real field  is done in the usual way, by first considering the full solution to the equations of motion and then imposing the real valuedness, as well as assuming the canonical commutation relations.
As a result of this, we  have the representation in terms of creation and annihilation operators,
\begin{equation}
\phi(x,t)=\int {[dk]\over \sqrt{2\omega(k)}} \left( a^\dagger(k)e^{-ikx+i\omega(k) t}+a(k)e^{ikx-i\omega(k)t}\right)
,\end{equation}
where
\begin{equation}
[a(k), a^\dagger(p)]=\delta[k-p]\quad {\rm and} \quad \omega(k)=\sqrt{k^2+m^2}
.\end{equation}
We have the nonlocal splitting into positive and negative frequency (energy) parts,
\begin{equation}
\phi^{(+)}(x,t)=\int {[dk]\over \sqrt{2\omega(k)}} a(k)e^{ikx-i\omega(k)t}
,\end{equation}
similarly for the negative frequency part. The free Hamiltonian part of the relativistic field itself does not 
evolve in time, as can be verified. Our choice of the interaction forbids the particle creation-annihilation process, it is not truely a delta function, it is a truncated version of it, in the limit where the particle creation-annihilation is negligible. 
As a first approximation to this still complicated problem, we will invoke a kind of Born-Oppenheimer approximation and pretend that the heavy particles are actually static, therefore are completely localized at $x_1$ and $x_2$. This approximation can be formulated in the following way,
\begin{equation}
H^{(0)}=\int [dk]\sqrt{k^2+m^2} a^\dagger(k)a(k)-\lambda \sum_{i=1,2}\phi^{(-)}(x_i)\phi^{(+)}(x_i) 
.\end{equation}
A nice way to attack this problem is to utilize a discrete version of the orthofermion algebra as it was done in \cite{caglar-teo}, as a result one finds the following $\Phi_{ij}(\mu)$ operator, where we wrote 
$-m<\mu(|x_1-x_2|)<m$ for the binding energy, while emphasizing its  dependence on the distance between the two centers:
\begin{equation}
\Phi(\mu(|z|))=\begin{cases} {1\over \lambda}-{1\over 2}\int_{-\infty}^\infty [dp]\frac{1}{(\sqrt{p^2+m^2})(\sqrt{p^2+m^2}-\mu(|z|))}& \mbox{if } i=j\\ -{1\over 2}\int_{-\infty}^\infty [dp] \frac{e^{ipz}}{\sqrt{p^2+m^2}(\sqrt{p^2+m^2}-\mu(|z|))}& \mbox{if } i\neq j \end{cases}
,\end{equation}
where we set $z=x_1-x_2$. To find the ground state energy we need to study the solutions of the  $zero$ eigenvalues of the matrix $\Phi_{ij}$, and similar to the nonrelativistic case, one finds that the symmetric solution corresponds to lower energy, which, in turn corresponds to the solution of the following equation:
\begin{equation}
\frac1{\lambda} -\int_{-\infty}^\infty {[dp]\over 2} \frac{1}{\sqrt{p^2+m^2}(\sqrt{p^2+m^2}-\mu(|z|))} =\int_{-\infty}^\infty {[dp]\over 2} \frac{e^{ipz}}{\sqrt{p^2+m^2}(\sqrt{p^2+m^2}-\mu(|z|))}
.\end{equation}
As we will see shortly, the binding energy decreases as we increase the distance in this case, hence the most strongly bound case, that is, the true ground state enery is found by setting $z=0$.
This corresponds to  the binding of the two heavy particles, a kind of  molecule formation. It is possible to get this zeroth order solution as,
\begin{equation}
\frac{1}{\lambda}-\int_{-\infty}^\infty {[dp]\over 2}\frac{1}{(\sqrt{p^2+m^2})(\sqrt{p^2+m^2}-\mu_0)}=\int_{-\infty}^\infty{[ dp]\over 2}\frac{1}{(\sqrt{p^2+m^2})(\sqrt{p^2+m^2}-\mu_0)}
.\end{equation}
Let us cast this expression into something that we can work with. To this purpose, we use a
Feynman parametrization,  
\begin{equation}
\int_{-\infty}^\infty [dp]\frac{1}{(\sqrt{p^2+m^2})(\sqrt{p^2+m^2}-\mu_0)} =\int_0^1 du\int_{-\infty}^\infty [dp]\frac{1}{(\sqrt{p^2+m^2}-\mu_0 u)^2}
,\end{equation}
then we employ an exponentiation trick, 
\begin{equation}
\frac{1}{(\sqrt{p^2+m^2}-\mu_0 u)^2} =\int_0^\infty dt\, t\, e^{-t(\sqrt{p^2+m^2}-\mu_0 u)}
,\end{equation}
and finally, 
to calculate the resulting  integral,  we resort  to the  subordination identity, given by
\begin{equation}
e^{-t\sqrt{p^2+m^2}}=\frac{1}{2\sqrt{\pi}}\int_0^\infty ds \frac{t}{s^{\frac3{2}}}e^{-s(p^2+m^2)-\frac{t^2}{4s}}
.\end{equation}
Collecting all these manipulations, we find,
\begin{equation}
\int_{-\infty}^\infty [dp] \frac{1}{\sqrt{p^2+m^2}(\sqrt{p^2+m^2}-\mu_0 )}= \frac1{2\sqrt{\pi}} \int_0^1 du\int_0^\infty dt t^2e^{ut\mu_0}\int_0^\infty ds\frac1{s^{\frac3{2}}}e^{-m^2s-\frac{t^2}{4s}}\int_{-\infty}^\infty [dp] e^{-sp^2},
\end{equation}
after performing the momentum integral we end up with,
\begin{equation}
\int_{-\infty}^\infty [dp] \frac{1}{\sqrt{p^2+m^2}(\sqrt{p^2+m^2}-\mu_0 )} =\frac{1}{4\pi}\int_0^1 du \int_0^\infty dt t^2e^{ut\mu_0}\int_0^\infty ds \frac{1}{s^2}e^{-sm^2-\frac{t^2}{4s}}.
\end{equation}
Now we recall the well-known formula for the modified Bessel functions (for most of the standart functions and integrals thereof we use \cite{gradsh}),
\begin{equation}
K_{\nu}(w) =\frac1{2}\left(\frac{w}{2}\right)^\nu\int_0^\infty ds \frac{e^{-s-\frac{w^2}{4s}}}{s^{\nu+1}}
.\end{equation}
Using this we simplify our expression into,
\begin{equation}
\int_{-\infty}^\infty [dp] \frac{1}{\sqrt{p^2+m^2}(\sqrt{p^2+m^2}-\mu_0 )}  ={m\over \pi}\int_0^1 du\int_0^\infty dt t e^{ut\mu_0}K_1(mt)= \frac{m}{\pi \mu_0}\int_0^\infty dt(e^{\mu_0 t}-1)K_1(mt).
\end{equation}
We employ the identity,
\begin{equation}
K_1(mt) =-\frac1{m}\frac{\partial K_0(mt)}{\partial t}
\end{equation}
to rewrite this expression as,
\begin{equation}
\int_0^\infty dt K_1(mt)(e^{\mu_0 t}-1) =\frac1{m} (1-e^{\mu_0 t})K_0(mt)|_0^\infty +\frac{\mu_0}{m}\int_0^\infty dt e^{\mu_0 t}K_0(mt).
\end{equation}
Note that, 
\begin{equation}
(e^{\mu_0 t}-1)K_0(mt)|_0^\infty = 0 
,\end{equation}
as long as $\mu_0<m$, 
so we can calculate the remaining integral, and find,
\begin{equation}
\int_0^\infty dte^{\mu_0 t}K_0(mt) =\frac{ \arccos(-\frac{\mu_0}{m})}{\sqrt{m^2-\mu_0^2}}
.\end{equation}
So finally, we have,
\begin{equation}
\int_{-\infty}^\infty [dp] \frac1{\sqrt{p^2+m^2}(\sqrt{p^2+m^2}-\mu_0 )} = {1\over \pi} \frac{ \arccos(-\frac{\mu_0}{m})}{\sqrt{m^2-\mu_0^2}},
\end{equation}
or inserting this expression back into the eigenvalue equation, we have for $\mu_0$ the equation,
\begin{equation}
\frac1{\lambda} ={1\over \pi}\frac{ \arccos(-\frac{\mu_0}{m})}{\sqrt{m^2-\mu_0^2}}
.\end{equation}
This equation will always have a solution for  any given  $\lambda>0$, up to  $\lambda\mapsto (\pi m)^{-}$ which  actually leads to $\mu_0\mapsto -m^+$. We may consider  $\lambda=2m$ as the critical value, since $\mu_0\mapsto 0^+$ corresponds to  this critical value.
Let us now obtain the effective potential generated by a small separation of the heavy particles, that is, we assume that 
  $ z\ne 0$, yet it is small, in a sense to be made more precise later on.
This corresponds to the idea that the light degrees of freedom is averaged out assuming some arbitrary (yet small)  $z$ in the background. Hence we should write an  effective Hamiltonian for this case again as in the nonrelativistic version, or equivalently an effective Schr\"odinger equation, 
\begin{equation}
\Big[-\frac{1}{2M}\sum_{i}{\nabla_i}^2+\mu(|x_1-x_2|)\Big]\psi(x_1,x_2)=E\psi(x_1,x_2)
.\end{equation} 
To find the dynamics of the heavy system we need the effective potential term in between. 
Assuming in our solution $|z|$ remains small, so we also get a small correction to $\mu_0$,  we will expand this term. In order to do this, we write the general case, 
\begin{equation}
\frac{1}{\lambda} -\int_{-\infty}^\infty {[dp]\over 2} \frac1{\sqrt{p^2+m^2}(\sqrt{p^2+m^2}-(\mu_0+\delta\mu))} =\int_{-\infty}^\infty {[dp]\over 2} \frac{e^{ipz}}{\sqrt{p^2+m^2}(\sqrt{p^2+m^2}-(\mu_0+\delta\mu(|z|)))},
\end{equation}
or we can write it as, 
\begin{equation}
\frac{1}{\lambda}-\frac{1}{2\pi}\frac{\arccos(-\frac{(\mu_0+\delta\mu)}{m})}{\sqrt{m^2-(\mu_0+\delta\mu)^2}} =\int_{-\infty}^\infty {[dp]\over 2} \frac{e^{ipz}}{\sqrt{p^2+m^2}(\sqrt{p^2+m^2}-(\mu_0+\delta\mu))}
.\end{equation}
By going through similar steps as before for the righthand side, we end up with,
\begin{eqnarray}
\int_{-\infty}^\infty [dp] \frac{e^{ipz}}{\sqrt{p^2+m^2}(\sqrt{p^2+m^2}-(\mu_0+\delta\mu))}&=&
 \frac{1}{4\pi}\int_0^1 du\int_0^\infty dt t^2e^{ut(\mu_0+\delta\mu)}\int_0^\infty ds\frac{e^{-m^2s-\frac{(t^2+z^2)}{4s}}}{s^2} \nonumber\\
&=&
\frac{m}{(\mu_0+\delta\mu)\pi}\int_0^\infty dtt \frac{(e^{t(\mu_0+\delta\mu)}-1)K_1(m\sqrt{t^2+z^2})}{\sqrt{t^2+z^2}}\nonumber
\end{eqnarray}
We now use the relationship between $K_1(Z)$ and $K_0(Z) $, we notice that, 
$$
\frac{tK_1(m\sqrt{t^2+z^2})}{\sqrt{t^2+z^2}}=-\frac1{m}\frac{\partial K_0(m\sqrt{t^2+z^2})}{\partial t}
$$
hence, 
\begin{equation}
\int_{-\infty}^\infty [dp]\frac{e^{ipz}}{\sqrt{p^2+m^2}(\sqrt{p^2+m^2}-(\mu_0+\delta\mu))}={1\over \pi}\int_0^\infty dte^{t(\mu_0+\delta\mu)}K_0(m\sqrt{t^2+z^2}).
\end{equation}
This finally leads to the result
\begin{equation}
\frac1{\lambda} -{1\over 2\pi}\frac{\arccos(-\frac{(\mu_0+\delta\mu)}{m})}{\sqrt{m^2-(\mu_0+\delta\mu)^2}}={1\over2\pi}\int_0^\infty dte^{t(\mu_0+\delta\mu)}K_0(m\sqrt{t^2+z^2})
\end{equation}
By expanding the energy equation in  $\delta \mu$,
\begin{equation}
\frac{1}{\lambda}-\frac1{2\pi}\frac{\arccos\left(-\frac{\mu_0}{m}\right)}{\sqrt{m^2-\mu_0^2}}-\frac{\delta\mu}{2\pi}\left(\mu_0\frac{\arccos\left(-\frac{\mu_0}{m}\right)}{(m^2-\mu_0^2)^\frac{3}{2}}+\frac1{m^2-\mu_0^2}\right)=\frac1{2\pi}\int_0^\infty dt e^{t(\mu_0+\delta\mu)}K_0(m\sqrt{t^2+z^2})
,\end{equation}
or equivalently,
\begin{equation}
\frac{1}{\lambda}-\frac{1}{\pi}\frac{\arccos\left(-\frac{\mu_0}{m}\right)}{\sqrt{m^2-\mu_0^2}}-\frac{\delta\mu}{m^2-\mu_0^2}\left({\mu_0\over 2\lambda}+{1\over 2\pi}\right)=\frac1{2\pi}\int_0^\infty dt [e^{t(\mu_0+\delta\mu)}K_0(m\sqrt{t^2+z^2})-e^{t\mu_0}K_0(mt)]
.\end{equation}
By also expanding the $\delta\mu$ term on the right side, we have,
\begin{equation}
\delta\mu\int_0^\infty dt te^{t\mu_0}K_0(m\sqrt{t^2+z^2}) \rightarrow \delta\mu \int_0^\infty dtte^{t\mu_0}K_0(mt)=\delta\mu\frac{\partial}{\partial\mu_0}\int_0^\infty dt e^{t\mu_0}K_0(mt).
\end{equation}
And we notice that 
\begin{equation}
\delta\mu\frac{\partial}{\partial\mu_0}\int_0^\infty dt e^{t\mu_0}K_0(mt)=\delta\mu\left[\mu_0\frac{\arccos\left(-\frac{\mu_0}{m}\right)}{(m^2-\mu_0^2)^\frac{3}{2}}+\frac1{m^2-\mu_0^2}\right]
.\end{equation}
As a result we end up with,
\begin{equation}
\frac{1}{\lambda}-\frac{1}{\pi}\frac{\arccos\left(-\frac{\mu_0}{m}\right)}{\sqrt{m^2-\mu_0^2}}-\frac{\delta\mu}{m^2-\mu_0^2}\left({\mu_0\over \lambda}+{1\over \pi}\right)=\frac1{2\pi}\int_0^\infty dt e^{t\mu_0}[K_0(m\sqrt{t^2+z^2})-K_0(mt)]
.\end{equation}
We recognize that the first two terms correpond to the equation defining $\mu_0$, moreover we may compute the left side by expanding in $z$, assuming the wave function corresponding to this potential keeps  $z$  small, we find,
\begin{equation}
-\frac{\delta\mu}{m^2-\mu_0^2}\left({\mu_0\over \lambda}+{1\over \pi}\right)=-{1\over 4}|z|
\end{equation}
or we find the first order effective potential term, to be used in the Schr\"odinger equation, as,
\begin{equation}
\delta\mu(z)=\frac{|z|}{4}\frac{(m^2-\mu_0^2)\lambda}{\left(\mu_0+\frac{1}{\pi}\lambda\right)}
.\end{equation}
We relegate the derivation of this term in  Appendix-II not to distract the reader from the main line of argument.
Thus, we arrive the Schr\"odinger equation, separating the center of mass motion again and using the wave function for the relative coordinate, reduced mass being equal to $M/2$, gives us,
\begin{equation}
\Big[-\frac{1}{M}{\partial^2\over \partial z^2}+\frac{|z|}{4}\frac{(m^2-\mu_0^2)\lambda}{\left(\mu_0+\frac{1}{\pi}\lambda\right)}\Big]\psi(z)=(E-{Q^2\over 4M})\psi(z)=\delta E \psi(z)
.\end{equation} 
The rest of the argument goes similar to the nonrelativistic case.

\section{Once Again A Many Body View }

In this section, we will approach the same problem directly from the many-body formulation, hoping that this will give us some more insight on the Born-Oppenheimer approximation. Let us write down the resulting principal operator $\Phi$ in this case, again introducing the orthofermion Fock space as in the nonrelativistic situation. For the sake of brevity, we skip details. We use momentum space description and use the fact that there is a single relativistic particle and two heavy particles, hence after normal ordering $\Phi$ acts on no-light particle, one orthofermion plus one heavy particle Fock state. This means we have,
\begin{eqnarray}
\Phi &=& \frac1{\lambda}\Pi_1 -\int [dp dq]\chi^{\dagger}(p+q)\frac1{2\sqrt{q^2+m^2}[\sqrt{q^2+m^2}+H_0+\frac{p^2}{2M}-(\mu_0+\delta' E)]}\chi(p+q)
\nonumber\\
&\ &-\int [dpdqdr]\chi^{\dagger}(p)\Psi^{\dagger}(r-q)\frac1{2\sqrt{q^2+m^2}[{r^2\over 2M}+{p^2\over 2M}+\sqrt{q^2+m^2}-(\mu_0+\delta' E)]}\Psi(p-q)\chi(r)\nonumber
,\end{eqnarray}
where we wrote for the energy, $E$,  an expansion $E=\mu_0+\delta' E$ anticipating higher order terms, and for the potential part we dropped $H_0$, here $H_0$ only refering to  the heavy particle free Hamiltonian, since there are no heavy particles left after normal ordering of the potential part, acting on the special sector that we are interested in.

Let us first understand the operator which only contains orthofermion creation and annihilation operators, by using Feynmann parametrization, exponentiation and subordination consecutively,
\begin{equation}
\int [dpdq]\chi^{\dagger}(p+q)\frac1{\sqrt{q^2+m^2}[\sqrt{q^2+m^2}+H_0+\frac{p^2}{2M}-(\mu_0+\delta' E)]}\chi(p+q)=\nonumber
\end{equation}
$$
\frac1{2\sqrt{\pi}}\int [dpdq]\chi^{\dagger}(p+q)\int_0^1 du\int_0^\infty dt t^2e^{-ut(H_0+\frac{p^2}{2M}-(\mu_0+\delta' E))}\int_0^\infty ds \frac{e^{-s(m^2+q^2)-\frac{t^2}{4s}}}{s^\frac3{2}}\chi(p+q)
.$$
Let us  define new coordinates   as 
$$ p+q= R$$and  $$p-\alpha q=Q$$ with some $\alpha$ and require that we can write the sum of the energy expressions as a sum of two squares,
\begin{equation}
\frac{tup^2}{2M}+sq^2 =A(p+q)^2+B(p-\alpha q)^2
.\end{equation}
This allows us to calculate these unkown constants $\alpha, A$ and $B$ as,
\begin{equation}
\alpha =\frac{2Ms}{tu}
,\quad
A= \frac{stu}{2M}\left(\frac1{\frac{tu}{2M}+s}\right)\quad {\rm and} \quad
B=\left(\frac{tu}{2M}\right)^2\left(\frac1{\frac{tu}{2M}+s}\right)
.\end{equation}
As a result the Jacobian of this transformation becomes, 
\begin{equation}
[dpdq]=\left(\frac{tu}{2M}\right)\left(\frac1{\frac{tu}{2M}+s}\right)[dRdQ]
.\end{equation}
Subsequently,  we can rewrite this part of the operator $\Phi$ again as,
\begin{eqnarray}
&\ &\!\!\!\!\!\!\!\!\int [dpdq]\chi^{\dagger}(p+q)
\frac1{\sqrt{q^2+m^2}[\sqrt{q^2+m^2}+H_0 +\frac{p^2}{2M} -(\mu_0+\delta' E)]}\chi(p+q)\nonumber\\
&\ &\quad =\frac1{2\sqrt{\pi}}\int [dR]\chi^{\dagger}(R)e^{-AR^2}\int_0^1 du \int_0^\infty dt\, t^2 e^{-ut[H_0-(\mu_0+\delta' E)]}\nonumber\\
&\ &\quad \quad \quad \quad \quad \quad \quad \quad \times\int_0^\infty ds\frac{e^{-sm^2-\frac{t^2}{4s}}}{s^{\frac3{2}}}\int [dQ]\left(\frac{tu}{2M}\right)\frac{e^{-BQ^2}}{\left(\frac{tu}{2M}+s\right)}\chi(R)
.\end{eqnarray}
One can evaluate $dQ$-integral  and this expression takes the form,
\begin{equation}\label{key1}
\frac{1}{4\pi}\int [dR]\chi^{\dagger}(R)e^{-AR^2}\int_0^1 du \int_0^\infty dt t^2 e^{-ut(H_0-(\mu_0+\delta' E))}\int_0^\infty ds \frac{e^{-sm^2-\frac{t^2}{4s}}}{s^{\frac3{2}}\sqrt{\frac{tu}{2M}+s}}\chi(R).
\end{equation}
Here $M$ is very large compared to $m$. Therefore we expect that some of the terms here are of lower order hence can be neglected. To be precise, $tu/M$ terms can be neglected in comparison to $s$ terms to leading order. To make sense of this claim  we need to scale $s$ by $1/m^2$ to get dimensionless variables, and also  $t$ by $1/m$. This argument is given in  Appendix-III, after neglecting this term, the result becomes, 
\begin{eqnarray}
&\ &\frac1{4\pi}\int [dR]\chi^{\dagger}(R)\int_0^1 du \int_0^\infty dt t^2e^{-ut(H_0+\frac{R^2}{2M}-(\mu_0+\delta'E))}\int_0^\infty ds\frac{e^{-sm^2-\frac{t^2}{4s}}}{s^2}\chi(R) \nonumber\\
&\ &\quad\quad  \quad =-{m\over\pi} \int [dR]\chi^{\dagger}(R)\int_0^\infty dt\frac{ (e^{-t(H_0+\frac{R^2}{2M}-(\mu_0+\delta' E))}-1)}{(H_0+\frac{R^2}{2M}-(\mu_0+\delta' E))}K_1(mt)\chi(R)\nonumber\\
&\ &\quad \quad \quad ={1\over \pi}\int [dR]\chi^{\dagger}(R)\frac{\arccos(\frac{H_0+\frac{R^2}{2M}-\delta' E-\mu_0}{m})}{\sqrt{m^2-(H_0+\frac{R^2}{2M}-\delta' E-\mu_0)^2}}\chi(R)
\end{eqnarray}	
Here, we changed  the first exponential term due to the replacement,
\begin{equation}
A =\left(\frac{stu}{2M}\right)\left(\frac1{\frac{tu}{2M}+s}\right)\rightarrow \frac{tu}{2M}
.\end{equation}
So finally, the operator $\Phi$ takes the form,
\begin{eqnarray}
\Phi=\frac1{\lambda}\Pi_1&-&\!\!\!\!\!\frac1{2\pi}\int [dR]\chi^\dagger(R)\frac{\arccos\left(\frac{\frac{R^2}{2M}+H_0-\mu_0-\delta' E}{m}\right)}{\sqrt{m^2-\left(\frac{R^2}{2M}+H_0-\mu_0-\delta' E\right)^2}}\chi(R)\nonumber\\
&-&\!\!\!\!\int [dpdqdr]\chi^\dagger(p)\Psi^\dagger(r-q)\frac1{2\sqrt{q^2+m^2}(\sqrt{q^2+m^2}-\mu_0 +{r^2\over 2M}+{p^2\over 2M}-\delta' E)}\Psi(p-q)\chi(r)\nonumber
\end{eqnarray}

We now note that the "kinetic" part of  $\Phi$  operator can be expanded, assuming all the kinetic energy contributions of the heavy degrees of freedom are small compared to the binding energy $m-\mu_0$, thus,
\begin{eqnarray}
&\ &\frac{\arccos\left(\frac{\frac{R^2}{2M}+H_0-\mu_0-\delta\mu}{m}\right)}{\sqrt{m^2-\left(\frac{R^2}{2M}+H_0-\mu_0-\delta' E\right)^2}}=\left(\arccos\left(-\frac{\mu_0}{m}\right)-\frac{H_0+\frac{R^2}{2M}-\delta' E}{\sqrt{m^2-\mu_0^2}}\right)\nonumber\\
&\ & \quad \quad\qquad \qquad \qquad\qquad\qquad \qquad \times\frac{1}{\sqrt{m^2-\mu_0^2}}\left(1-\mu_0\frac{(H_0+\frac{R^2}{2M}-\delta' E)}{m^2-\mu_0^2}\right)+...\nonumber
\end{eqnarray}
We may now further expand the small part of the potential term, 
\begin{eqnarray}
&\ &\int [dpdqdr]\chi^\dagger(p)\Psi^\dagger(r-q)\frac1{2\sqrt{q^2+m^2}(\sqrt{q^2+m^2}-\mu_0 +{r^2\over 2M}+{p^2\over 2M}-\delta' E)}\Psi(p-q)\chi(r)\nonumber\\
&\ &=\int [dpdqdr]\chi^\dagger(p)\Psi^\dagger(r-q)\frac1{2\sqrt{q^2+m^2}(\sqrt{q^2+m^2}-\mu_0 )}\Psi(p-q)\chi(r)\nonumber\\
&\ &-\int [dpdqdr]\chi^\dagger(p)\Psi^\dagger(r-q)\frac{({r^2\over 2M}+{p^2\over 2M}-\delta' E)}{2\sqrt{q^2+m^2}(\sqrt{q^2+m^2}-\mu_0 )^2}\Psi(p-q)\chi(r)+...
\end{eqnarray}
This leads to an expansion of $\Phi(E)$,
\begin{eqnarray}
\Phi&=&\!\!\!\left[\frac1{\lambda}-\frac1{2\pi}\frac{\arccos\left(-\frac{\mu_0}{m}\right)}{\sqrt{m^2-\mu_0^2}}\right]\Pi_1\nonumber\\
&+&\!\!\!\frac1{2\pi}\left(\mu_0\frac{\arccos\left(-\frac{\mu_0}{m}\right)}{(m^2-\mu_0^2)^\frac{3}{2}}+\frac1{m^2-\mu_0^2}\right)\int [dR]\chi^\dagger(R)\left(H_0+\frac{R^2}{2M}-\delta'E\right)\chi(R)\nonumber\\
&-&\!\!\!{1\over 2}\int [dpdqdr]\chi^\dagger(p)\Psi^\dagger(r-q)\frac1{\sqrt{q^2+m^2}(\sqrt{q^2+m^2}-\mu_0)}\Psi(p-q)\chi(r)\nonumber\\
&+&\!\!\!{1\over 2}\int [dpdqdr]\chi^\dagger(p)\Psi^\dagger(r-q)\frac{({r^2\over 2M}+{p^2\over 2M}-\delta' E)}{\sqrt{q^2+m^2}(\sqrt{q^2+m^2}-\mu_0 )^2}\Psi(p-q)\chi(r)+...\nonumber
.\end{eqnarray}
Let's define a particular  wave function as our ansatz for the solution,
$$
|\omega>=\int [dQd\xi]e^{-iQX}f(\xi)\chi^\dagger(Q/2+\xi)\Psi^\dagger(Q/2-\xi)|\Omega>
,$$
and demand as before 
\begin{equation}
\Phi(\mu_0+\delta' E) |\omega>=0
.\end{equation}
When we solve this equation order by order, we will find $f$ and the true energy of the system. Going through the same discussion as in the nonrelativistic case this problem can be reduced to an eigenfunction and eigenvalue equation for $f$. Again we assume $f$ is symmetric, and the last part of the potential term again can be turned into a convolution in two different ways, which corresponds to the ordering ambiguity. We will spare the details since they are very similar to the previous case. The only thing we should find is the corrected potential corresponding to the inverse Fourier transform of 
\begin{equation}
\frac{1}{\sqrt{q^2+m^2}(\sqrt{q^2+m^2}-\mu_0 )^2}
.\end{equation}
Going through exactly the same steps as before, using a Feynmann parametrization and an exponentiation, recognizing the modified Bessel function inside, 
we obtain the inverse Fourier transform as,
\begin{eqnarray}
\int_{-\infty}^\infty [dp] \frac{e^{ipz}}{\sqrt{p^2+m^2}(\sqrt{p^2+m^2}-\mu_0)^2}&=&
 \frac{1}{4\pi}\int_0^1 u du\int_0^\infty dt t^3e^{ut\mu_0}\int_0^\infty ds\frac{e^{-m^2s-\frac{(t^2+z^2)}{4s}}}{s^2} \nonumber\\
&=&
\frac{m}{\pi\mu_0}\int_0^1 u du\int_0^\infty dtt^3 \frac{e^{ut\mu_0}K_1(m\sqrt{t^2+z^2})}{\sqrt{t^2+z^2}}\nonumber\\
&=&\frac{m}{\pi \mu_0}{\partial \over \partial \mu_0}\int_0^1  du\int_0^\infty dtt^2 \frac{e^{ut\mu_0}K_1(m\sqrt{t^2+z^2})}{\sqrt{t^2+z^2}}\nonumber\\
&=&{1 \over \pi}{\partial\over \partial \mu_0}\int_0^\infty dt e^{\mu_0 t} K_0(m\sqrt{t^2+z^2})\nonumber
\end{eqnarray}
Inserting all these expressions back, we end up with an eigenvalue equation, 
\begin{eqnarray}
&\ & \left(\frac1{\lambda}-\frac1{2\pi}\frac{\arccos\left(-\frac{\mu_0}{m}\right)}{\sqrt{m^2-\mu_0^2}}\right)f(z)+\frac1{2\pi}\left(\mu_0\frac{\arccos\left(-\frac{\mu_0}{m}\right)}{(m^2-\mu_0^2)^\frac{3}{2}}+\frac1{m^2-\mu_0^2}\right)\left(-\frac{\nabla^2_z}{M}+\frac1{2}\frac{Q^2}{2M}-\delta' E\right)f(z)\nonumber\\
&\ & -{1\over 2\pi}\left(\int_0^\infty dt e^{\mu_0 t} K_0(m\sqrt{t^2+z^2} )\right) f(z)\nonumber\\
&\ & +{1\over 2}\left[-\frac{\nabla^2_z}{M}+\frac1{2}\frac{Q^2}{2M}-\delta' E \, , {1\over 2\pi}{\partial\over \partial \mu_0}\int_0^\infty dt e^{\mu_0 t} K_0(m\sqrt{t^2+z^2})\right]_+ f(z)=0\nonumber
\end{eqnarray}
We now introduce the small parameter $\delta E=\delta' E-{Q^2\over 4M}$, and expanding the third  potential term up to $|z|$, and keeping only the constant contribution from the last anticommutator, dropping all the derivative corrections to this order, we have,
\begin{eqnarray}
&\ & \left(\frac1{\lambda}-\frac{1}{\pi}\frac{\arccos\left(-\frac{\mu_0}{m}\right)}{\sqrt{m^2-\mu_0^2}}\right)f(z)\nonumber\\
&\ & \ \ \  +\frac{1}{\pi}\left(\mu_0\frac{\arccos\left(-\frac{\mu_0}{m}\right)}{(m^2-\mu_0^2)^\frac{3}{2}}+\frac1{m^2-\mu_0^2}\right)\left(-\frac{\nabla^2_z}{M}-\delta E\right)f(z)
 +{1\over 4} |z|f(z)=0\nonumber
.\end{eqnarray}
If we now look at the zeroth order solution, we find as expected the equation for $\mu_0$ as before, but that does not determine $f$.
 Assuming $\mu_0$ solves the zeroth order equation, thus removing the constant multiple of $f$ in the equation, the next order equation becomes,
$$
\left(-\frac{\nabla^2_z}{M}+\frac1{2}\frac{Q^2}{2M}-\delta E\right)f(z)+\frac1{4}\frac{(m^2-\mu_0^2)}{(\mu_0+\frac{1}{\pi}\lambda)}\lambda|z|f(z)=0
,$$
which is exactly as before.
Nevertheless, we remark that this picture again allows us to go one step further and obtain the second order corrections to the energy.
We leave this more cumbersome calculation to the reader, which can be found exactly following the previous part.

\section{Conclusions} 

We show that for a simple model which consists of two heavy particles interacting with a light particle through attractive delta function potentials in one dimension, we can apply the Born-Oppenheimer approximation to understand the spectrum. The simplicity of the model allows us to test the validity of the Born-Oppenheimer approximation self-consistently. A novel approach, originally proposed by Rajeev for interacting bosons in two dimensions can also be applied  to this problem not only to recover the previous results but also obtain in a more systematic way higher order corrections. It turns out that for this system the relevant expansion paramater  becomes $(m/M)^{1/3}$. A modification of this problem where the light particle is very small   yet the binding is not so strong to cause pair creation is also proposed. Along similar lines this problem is discussed. The many body perspective can be adapted to this problem as well and we can again recover the Born-Oppenheimer result from this point of view. 

\section{Appendix-I}

In this appendix we  present the error estimates for the Born-Oppenheimer Approximation.
In order to do this we first write down the 
the wave function normalization for the light degree of freedom.
Let us recall that the wave function we found is given by,
\begin{equation}
\phi(x|x_1,x_2) =\frac{N}{\hbar}\left[\int_0^\infty dt K_t(x,x_1)e^{-\frac{\nu^2}{\hbar}t}+\int_0^\infty dt K_t(x,x_2)e^{-\frac{\nu^2}{\hbar}t}\right]
.\end{equation}
We demand the normalization,
$$
\int_{-\infty}^\infty dx |\phi(x|x_1,x_2)|^2= 1
$$
 The convolution property of the heat kernel, 
\begin{equation}
\int dx K_{t_1}(x,x_1)K_{t_2}(x,x_2)=K_{t_1+t_2}(x_1,x_2)
\end{equation}
simplifies the calculations, giving us the condition,
\begin{equation}
1=2\frac{N^2}{\hbar^2}\left[\int dt_1 dt_2 K_{t_1+t_2}(x_1,x_1)e^{-\frac{\nu^2}{\hbar} (t_1+t_2)}+\int dt_1 dt_2K_{t_1+t_2}(x_1,x_2)e^{-\frac{\nu^2}{\hbar}(t_1+t_2)} \right]
.\end{equation}
Defining new parameters,
\begin{equation}
t=t_1+t_2,  \quad
{\rm and} \quad
s=t_1-t_2
,\end{equation}
and executing the $s$-integrals,
$$
1=\frac{N^2}{\hbar^2}\left[\int_0^\infty dt\, t  [K_t(x_1,x_1)+K_t(x_2,x_2)]e^{-\frac{\nu^2}{\hbar}t}+2\int_0^\infty dt\, t  K_t(x_1,x_2)e^{-\frac{\nu^2}{\hbar}t}\right].
$$
The integrals can be calculated giving us the condition,
$$
1=\frac{N^2}{2}\frac{\sqrt{2m}}{\hbar\nu^3}\left[1+\left[1+\frac{\sqrt{2m}}{\hbar}\nu|x_1-x_2|\right]e^{-\frac{\sqrt{2m}}{\hbar}\nu|x_1-x_2|}\right].
$$
As a result the normalized light particle wave function becomes,
\begin{equation}
\phi(x|x_1,x_2) =\frac{\sqrt{\nu}}{\sqrt{2}} \frac{(2m)^\frac1{4}}{\sqrt{\hbar}}\frac1{\sqrt{\left[1+\left[1+\frac{\sqrt{2m}}{\hbar}\nu|x_1-x_2|\right]e^{-\frac{\sqrt{2m}}{\hbar}\nu|x_1-x_2|}\right]}}\left[e^{-\frac{\sqrt{2m}}{\hbar}\nu|x-x_1|}+e^{-\frac{\sqrt{2m}}{\hbar}\nu|x-x_2|}\right]
.\end{equation}
 Since the whole system is translationally invariant, it is more natural to define new coordinates, 
\begin{equation}
X=\frac{x_1+x_2}{2}\quad 
{\rm and }\quad 
z=x_1-x_2
\end{equation}
and express the wave function in terms of these coordinates, moreover, to isolate the relevant contributions, we shift all the coordinates by $X$, which will not change $z$, yet will modify $x$ to $x+X$, this gives us,
\begin{equation}
\phi(x,z)=\frac1{\sqrt{2}} \frac{(2m)^\frac1{4}}{\sqrt{\hbar}}\sqrt{\nu}\frac1{\sqrt{\left[1+\left[1+\frac{\sqrt{2m}}{\hbar}\nu|z|\right]e^{-\frac{\sqrt{2m}}{\hbar}\nu|z|}\right]}}\left[e^{-\frac{\sqrt{2m}}{\hbar}\nu|x-\frac{z}{2}|}+e^{-\frac{\sqrt{2m}}{\hbar}\nu|x+\frac{z}{2}|}\right]
.\end{equation}
In general, the derivatives with respect to $X$ coordinates may give us a large contribution, but that is the total translational kinetic energy of the system, which is not interesting from our perspective. Therefore, this shift of coordinates
will be natural in order  to remove such derivative contributions from our estimates. In principle, they can also be estimated but the results becomes  more combersome. 
Note that $\nu$ itself is  a function of $z$, as we determined  thorugh the eigenvalue equation previously,
\begin{equation}
{1\over \lambda} =\sqrt{m\over 2\hbar^2 \nu^2(z)}\Big(e^{-\sqrt{2m}\nu(z)/\hbar}+1\Big)
.\end{equation}
To simplify our estimates we rewrite the wave function in the following decomposed form,
\begin{equation}
\phi(z;x)= A(z)(\epsilon_+(x,z)+\epsilon_-(x,z)),
\end{equation}
where, 
$$
\epsilon_+= e^{-\frac{\sqrt{2m}}{\hbar}\nu|x+\frac{z}{2}|}\quad{\rm and} \quad
\epsilon_-= e^{-\frac{\sqrt{2m}}{\hbar}\nu|x-\frac{z}{2}|}
 $$
and $A(z)$ refers to the common  multiplicative part. Let us recall that within the approximations we use, $\nu(z)$ can be factored as, $\nu^2(z)=\nu^2_0+\delta E_1(z)$, with
$$
\nu_0=\frac{\sqrt{2m}}{\hbar}\lambda \quad {\rm and} \quad
\delta E_1=-{|z|}\lambda^3 \Big(\frac{2m}{\hbar^2}\Big)^2
.$$
Let us expand the derivatives acting on the product wave function as,
\begin{equation}
\left(\frac{\partial^2}{\partial x_1^2}+\frac{\partial^2}{\partial x_2^2}\right)\phi(x,z)\psi(z)=\psi(z)\left(\frac{\partial^2}{\partial x_1^2}+\frac{\partial^2}{\partial x_2^2}\right)\phi(x,z)+\phi(x,z)\left(\frac{\partial^2}{\partial x_1^2}+\frac{\partial^2}{\partial x_2^2}\right)\psi(z)
\end{equation}
$$
+2\frac{\partial \psi}{\partial x_1}\frac{\partial\phi}{\partial x_1}+2\frac{\partial \psi}{\partial x_2}\frac{\partial\phi}{\partial x_2}
.$$
We  use only the following  terms
$$
\psi(z)\left(\frac{\partial^2}{\partial x_1^2}+\frac{\partial^2}{\partial x_2^2}\right)\phi(x,z)+2\frac{\partial \psi}{\partial x_1}\frac{\partial\phi}{\partial x_1}+2\frac{\partial \psi}{\partial x_2}\frac{\partial\phi}{\partial x_2}
$$
since the derivatives acting on the wave function $\psi(z)$ is precisely the heavy particle part that we use in our approximations. Let us note that,
$$
\frac{\partial}{\partial x_1}=\frac1{2}\frac{\partial}{\partial X}+\frac{\partial}{\partial z}\quad {\rm and} \quad 
\frac{\partial}{\partial x_2}=\frac1{2}\frac{\partial}{\partial X}-\frac{\partial}{\partial z}
,$$
thus we need to estimate these two expressions,
\begin{equation}
2\psi(z)\frac{\partial^2}{\partial z^2}\phi(x,z)\quad {\rm and} \quad 4\frac{\partial\psi}{\partial z}\frac{\partial\phi}{\partial z}.
\end{equation}
We note that 
$$
\frac{\partial \epsilon_+}{\partial z}=\frac{1}{2}\frac{\partial\epsilon_+}{\partial x}-\frac{\sqrt{2m}}{\hbar}{\partial \nu\over \partial z} |x+\frac{z}{2}|\epsilon_+,
$$
$$
\frac{\partial\epsilon_-}{\partial z}=-\frac{1}{2}\frac{\partial\epsilon_-}{\partial x}-\frac{\sqrt{2m}}{\hbar}{\partial \nu\over \partial z} |x-\frac{z}{2}|\epsilon_-.
$$
We show the consistency of our approximations by evaluating the average of all these terms with the presumed solutions of the heavy particles wave functions $\psi$. 
We start with the second term,
$$
4\frac{\partial \phi}{\partial z}\frac{\partial\psi}{\partial z}=4\frac{\partial\psi}{\partial z}\left[\frac{\partial A}{\partial z}(\epsilon_++\epsilon_-)+A\left(\frac{1}{2}[\frac{\partial\epsilon_+}{\partial x}-\frac{\partial\epsilon_-}{\partial x}]-\frac{\sqrt{2m}}{\hbar}{\partial \nu\over \partial z}( |x-\frac{z}{2}|\epsilon_-+|x+\frac{z}{2}|\epsilon_+)\right)\right]
.$$
By taking the average of this term we estimate the contribution of this term we to the first order energy  calculation we already have. We replace this derivative term in the Schr\"odinger equation by the expectation value, 
\begin{eqnarray}
-\frac{4\hbar^2}{2\mu}\int dx dz \phi \psi \frac{\partial \phi}{\partial z}\frac{\partial\psi}{\partial z}&=&\!\!
-\frac{\hbar^2}{\mu}\Bigg[\int dz\psi\frac{\partial\psi}{\partial z}\int dx \left[2\frac{\partial A}{\partial z}(\epsilon_++\epsilon_-)+A\left(\frac{\partial\epsilon_+}{\partial x}-\frac{\partial\epsilon_-}{\partial x}\right)\right]A(\epsilon_++\epsilon_-)\nonumber\\
&-&\frac{\sqrt{2m}}{\hbar}{\partial \nu\over \partial z}( |x-\frac{z}{2}|\epsilon_-+|x+\frac{z}{2}|\epsilon_+) A^2(\epsilon_-+\epsilon_+)\Bigg]\nonumber\\
&=&-\frac{\hbar^2}{2\mu}\int dz\psi\frac{\partial \psi}{\partial z}\int dx \Bigg[4A\frac{\partial A}{\partial z}(\epsilon_++\epsilon_-)^2+2A^2\left(\epsilon_+\frac{\partial\epsilon_+}{\partial x}-\epsilon_-\frac{\partial\epsilon_-}{\partial x}\right)\nonumber\\
&\ &+2A^2\epsilon_+\epsilon_-\left(\frac1{\epsilon_+}\frac{\partial\epsilon_+}{\partial x}-\frac1{\epsilon_-}\frac{\partial \epsilon_-}{\partial x}\right)\nonumber\\
&\ &-2 A^2\frac{\sqrt{2m}}{\hbar}{\partial \nu\over \partial z}( |x-\frac{z}{2}|\epsilon_-+|x+\frac{z}{2}|\epsilon_+) (\epsilon_-+\epsilon_+)\Bigg]\nonumber
.\end{eqnarray}
To this purpose, we can  use the first order result for the derivative of energy,
\begin{equation}
{\partial \nu\over \partial z}\approx -{1\over 2}{\rm sgn} (z) \Big({2m\over \hbar^2} \Big)^{3/2} \lambda^2 
\end{equation}
Let us now look at the first term on the right side inside the integral, after $x$ integration it becomes,
\begin{equation}
(1)=-\frac{\hbar^2}{2\mu}4\int dz \psi \frac{\partial \psi}{\partial z}\frac{\partial}{\partial z}\ln A
.\end{equation}
The second term is an exact differential in $x$ and hence gives zero upon integration. The third term can be written as,
$$
(3)=-2\frac{\hbar^2}{2\mu}\int dxdz \psi\frac{\partial\psi}{\partial z}A^2\epsilon_+\epsilon_-\frac{\partial}{\partial x}\ln\left(\frac{\epsilon_+}{\epsilon_-}\right)
.$$
Last part contains various cross terms which we will estimate later.
Note that for the first  term it suffices to use the leading nonzero term of $\ln A$'s derivative,
$$
\frac{\partial}{\partial z}\ln A \approx \frac{\partial }{\partial z}\ln\sqrt{\nu}\approx -\frac1{4}\lambda\frac{2m}{\hbar^2}{\rm sgn}(z).
$$
As a result it has an upper bound,
$$\Bigg|\frac{\hbar^2}{2\mu}\nu_0\frac{\sqrt{2m}}{\hbar}\int dz {\rm sgn}(z)\psi \frac{\partial\psi}{\partial z}\Bigg|\le \nu_0\sqrt{\frac{m}{\mu}}\left|\int dz \left|\frac{\hbar}{\sqrt{2\mu}}\frac{\partial\psi}{\partial z}\right|^2\right|^\frac1{2}\left|\int dz |{\rm sgn}(z)\psi|^2\right|^\frac{1}{2}\le \nu_0^2\left(\frac{m}{\mu}\right)^\frac{2}{3}
,$$
where we used the fact that the kinetic energy is less than the total energy, since the potential, being proportional to $|z|$, is positive everywhere.
Let us come to the second nonzero term we found, 
$$
2\Bigg|\frac{\hbar^2}{2\mu}\frac{\sqrt{2m}}{\hbar}\nu_0\int dz \psi\frac{\partial\psi}{\partial z} A^2\int_{-\frac{|z|}{2}}^\frac{|z|}{2} dx e^{-\frac{\sqrt{2m}}{\hbar}\nu(|x+\frac{z}{2}|+|x-\frac{z}{2}|)}\frac{\partial}{\partial x}\left(|x+\frac{z}{2}|-|x-\frac{z}{2}|\right)\Bigg|
$$
$$
\le 16\frac{2m}{\hbar\sqrt{2\mu}}\nu_0^2\left|\int dz \left|\frac{\hbar}{\sqrt{2\mu}}\frac{\partial\psi}{\partial z}\right|^2\right|^\frac1{2}\left|\int dz  z^2|\psi|^2\right|^\frac{1}{2}\approx \frac{2m}{\hbar\sqrt{2\mu}}\nu_0^3\left(\frac{m}{\mu}\right)^\frac1{6}\sqrt{<z^2>}\approx C\frac{m}{\mu}\nu_0^2
$$
here we restrict the range of $x$ integration to $[-{|z|\over 2},{|z|\over 2}]$ because outside of this interval the expression inside the derivative has no dependence on $x$ hence gives zero upon differentiation. Since we are in a finite interval the exponential term is replaced with its upper bound  $1$, for this the absolute value of the derivative term ${\partial \psi\over \partial z}$ should be taken. Moreover, $A^2$ is replaced with its leading order constant value, since there is already $|z|$ term multiplying the whole expression as well as ${\partial \psi\over \partial z}$ term.

This leaves us with the last term, which contains various combinations like
\begin{equation}
 \int dx\, \epsilon_-\epsilon_+|x-{z\over 2}|
.\end{equation}
We note that the absolute value of all such combinations are smaller than the  following integral
\begin{equation}
2\int dx e^{-{\sqrt{2m} \over \hbar}\nu|x|} (|x|+|z|)
\end{equation}
as can be seen by using the $x\to -x$ transformation for the negative part of the real axis, replacing one of the exponentials by its upper limit $1$, afterwards shifting the integration variable, and 
then extending again the integration region to the full real axis.
This can be used to estimate the full expression,
\begin{equation}
(4)\leq 8 {\hbar^2\over \mu} {\sqrt{2m}\over \hbar} \int dz A^2 \Bigg| {\partial \nu\over \partial z}{\partial \psi\over \partial z}\Bigg| \psi \int dx  e^{-{\sqrt{2m} \over \hbar}\nu|x|} (|x|+|z|)
.\end{equation}
We will only estimate the $|x|$ term since it is easy to see that the $|z|$ term is even smaller.
Note that the $|x|$ part satisfies, to leading order,
\begin{equation}
(4')\leq 8 {\hbar\over \sqrt{\mu}} \Big({\sqrt{2m}\over \hbar}\Big)^{5}\nu_0 \Big({\sqrt{2m}\over \hbar}\nu_0\Big)^{-2}\lambda^2 \Big[ \int dz ({\rm sgn}(z)\psi)^2 \Big]^{1/2}  \Big[ {\hbar^2\over 2\mu}\int dz \Big|{\partial \psi\over \partial z}\Big|^2\Big]^{1/2}  
.\end{equation}
Here we replaced  $A^2$ by its constant value to leading order, that means $\nu$ is replaced  by $\nu_0$ to leading order everywhere.
Using $ {2m\over \hbar^2} \lambda^2=\nu_0^2$, we can reorganize these terms to find,
\begin{equation}
  (4')\leq C_4 \Big({m\over \mu}\Big)^{2/3} \nu_0^2
\end{equation}
as desired.

Let us now evaluate 
 the average of the second order derivatives,  $ [2\psi\frac{\partial^2}{\partial z^2}\phi]$, we divide this into a group of terms, since it is a long expression, we have the first group of terms, most conveniently written as averages,
$$
(1)={\hbar^2\over \mu} \int dx dz \psi^2 A (\epsilon_++\epsilon_-)\left[\frac{\partial^2A}{\partial z^2} (\epsilon_++\epsilon_-)+4\frac{\partial A}{\partial z}\left(\frac{\partial\epsilon_+}{\partial x}-\frac{\partial\epsilon_-}{\partial x}\right)+4A\left(\frac{\partial^2\epsilon_+}{\partial x^2}+\frac{\partial^2\epsilon_-}{\partial x^2}\right)\right]
.$$
Then we have a second group of terms after  taking into account the repetitions, we also face with second order derivatives. These are hard to estimate when we have only first order expansions given by simple expressions. To overcome this before we do any approximations, we first integrate by parts, to reduce these more complicated terms into products of derivatives.  As a result, there are  some terms {\it coming from the integration by parts inside the average}, all of it put together amount to the following,
\begin{eqnarray}
(2)={\hbar^2\over \mu } \int dx dz \psi^2 A(\epsilon_++\epsilon_-){\sqrt{2m}\over \hbar} {\partial A\over \partial z} {\partial \nu\over \partial z} \Big[\epsilon_+|x+{z\over 2} |+\epsilon_-|x-{z\over 2}|\Big]
\end{eqnarray}
Similarly, we use {\it an integration by parts trick} to write an average of a group of terms
\begin{eqnarray}
(3)&=&{\hbar^2\over \mu } \int dx dz \psi^2 A^2\Big[({\partial \epsilon_+\over \partial x}-{\partial \epsilon_-\over \partial x})-{\sqrt{2m}\over \hbar} {\partial \nu\over \partial z}(\epsilon_+|x+{z\over 2} |+\epsilon_-|x-{z\over 2}|)\Big]\nonumber\\
&\ & \quad \quad \quad \quad \quad\quad \quad \quad \quad \quad \ \ \ \ \ \ \ \ \ \ \  \times {\sqrt{2m}\over \hbar} {\partial \nu\over \partial z} \Big[\epsilon_+|x+{z\over 2} |+\epsilon_-|x-{z\over 2}|\Big]
.\nonumber\end{eqnarray}
Integration by parts in the average produces one more term of the form
\begin{equation}
\int dx dz A^2 (\epsilon_++\epsilon_-) \psi {\partial \psi\over \partial z} {\sqrt{2m}\over \hbar}  {\partial \nu\over \partial z} \Big[\epsilon_+|x+{z\over 2} |+\epsilon_-|x-{z\over 2}|\Big]
,\end{equation}
however this is exactly one of the terms we estimated on the cross-derivative terms above.
Let us now estimate the averages of the first group of terms,
$$
(1^a)=-2\frac{\hbar^2}{2\mu}\int dz |\psi|^2\frac1{A}\frac{\partial^2}{\partial z^2}A
$$
$$
(1^b)=-4\frac{\hbar^2}{2\mu}\int dxdz|\psi|^2\Big( \frac{\partial}{\partial z}A^2\Big)\Big(\epsilon_+\epsilon_-\frac{\partial}{\partial x}\ln\left(\frac{\epsilon_+}{\epsilon_-}\right)+{1\over 2}{\partial \over \partial x}( \epsilon_+^2+\epsilon_-^2 )
\Big)
$$
$$
(1^c)=-8\frac{\hbar^2}{2\mu}\int dxdz |\psi|^2 A^2\left(\frac{\partial^2\epsilon_+}{\partial x^2}+\frac{\partial^2\epsilon_-}{\partial x^2}\right)(\epsilon_++\epsilon_-)
.$$
Since it is hard to estimate the second derivative terms for normalization part, we note the identity, 
$$
\frac1{A}\frac{\partial^2}{\partial z^2}A =\frac{\partial^2}{\partial z^2}\ln A+\left(\frac{\partial}{\partial z}\ln A\right)^2
$$
Let us now consider 
the first integral expression $(1^a)$,
$$
-2\frac{\hbar^2}{2\mu}\int dz|\psi|^2\frac{\partial^2}{\partial z^2}\ln A=4\frac{\hbar^2}{2\mu}\int dz \psi\frac{\partial\psi}{\partial z}\frac{\partial\ln A}{\partial z}
$$
$$
\le\sqrt{\frac{m}{\mu}}\nu_0\left|\int dz\left|\frac{\hbar}{\sqrt{2\mu}}\frac{\partial\psi}{\partial z}\right|^2\right|^\frac1{2} \left|\int dz |\psi|^2\right|^\frac1{2}\approx \left(\frac{m}{\mu}\right)^\frac{2}{3}\nu_0^2
.$$
Moreover, the second term becomes,
$$
-2\frac{\hbar^2}{2\mu}\int dz |\psi|^2\left(\frac{\partial}{\partial z}\ln A\right)^2\approx\frac{m}{\mu}\nu_0^2.
$$
The second part $(1^b)$  has the following combination,
\begin{eqnarray}
4\frac{\hbar^2}{2\mu}\Bigg|\int dz|\psi|^2\Big( \frac{\partial}{\partial z}A^2\Big)\int^{|z|/2}_{-|z|/2} dx \epsilon_+\epsilon_-\frac{\partial}{\partial x}\ln\left(\frac{\epsilon_+}{\epsilon_-}\right)\Bigg|\le 8\frac{\hbar^2}{2\mu}\frac{(2m)^\frac{3}{2}}{\hbar^3}\nu_0^3\left[\int dz |z| |\psi|^2\right]\nonumber
,\end{eqnarray}
using $<|z|>=C {\lambda\over\nu_0^2}({m\over \mu})^{1/3}$, the above expression becomes,
$$
\le 8\frac{\hbar^2}{2\mu}\frac{(2m)^\frac{3}{2}}{\hbar^3}C {\lambda\over\nu_0^2}({m\over \mu})^{1/3}\nu_0^3 \approx \left(\frac{m}{\mu}\right)^\frac{4}{3}\nu_0^2.
$$
The expression for $(1^b)$ contains one more term of the form,
\begin{equation}
\int dx \Big( \epsilon_+{\partial \epsilon_+\over \partial x}+ \epsilon_-{\partial \epsilon_-\over \partial x}\Big)={1\over 2}\int dx {\partial \over \partial x}( \epsilon_+^2+\epsilon_-^2 ),
\end{equation}
which is a total derivative and integrates out to zero.
Let us now consider 
the last term, 
$$
(1^c)=-8\frac{\hbar^2}{2\mu}\int dxdz |\psi|^2 A^2\left(\frac{\partial^2\epsilon_+}{\partial x^2}+\frac{\partial^2\epsilon_-}{\partial x^2}\right)(\epsilon_++\epsilon_-)\approx C_8 \frac{m}{\mu}\nu_0^2,
$$
since
this integral is the average kinetic energy of the light particle  multiplied with $\frac{m}{\mu}$.
The potential energy can be computed using the known wave  function and shown to be less than 
some multiple of $\nu^2$ easily, this gives the estimate on the kinetic energy.
Due to the factor in front we can replace it with its leading constant value.

Let us discuss the terms in $(2)$,
we have combinations of the form, 
\begin{equation}
   \int dx \epsilon_{\pm} \epsilon_\pm |x\pm {z\over 2}|
\end{equation}
using our previous estimate on these terms, combining $A{\partial A\over \partial z}$ into ${1\over 2} {\partial A^2\over \partial z}$, and using the leading order term for this derivative, which is  given by ${\sqrt{2m}\over \hbar}{\partial \nu\over \partial z}$, and as we indicated before,   the leading term for ${\partial \nu\over \partial z}\approx -{1\over 2}$sgn$(z)({\sqrt{2m}\over \hbar})^3\lambda^2$, combining all these estimates and identities,  we find,
\begin{equation}
 | (2)| \leq C'_9 {\hbar^2\over \mu} \int dz \psi^2\Big( {\sqrt{2m}\over \hbar}\Big)^{2} \Big[ \Big({\sqrt{2m}\over \hbar}\Big)^3\lambda^2\Big]^2 \Big({\sqrt{2m}\over \hbar}\nu\Big)^{-2}\leq C_9 {m\over \mu} \nu_0^2
.\end{equation}

Let us consider the first group of terms in our combination labeled as $(3)$, integration by parts in $x$ leads to 
\begin{eqnarray}
\int dx dz \psi^2 A^2 (\epsilon_+-\epsilon_-){\sqrt{2m}\over \hbar}{\partial \nu\over \partial z} {\partial \over \partial x} (\epsilon_+|x+{z\over 2} |+\epsilon_-|x-{z\over 2}|)
.\end{eqnarray}
The last term in fact can be turned into $\epsilon_+|x+{z\over 2}|$ by using $z\to-z$ symmetry, and after that we can estimate the absolute value of each term.
Using, $|\epsilon_+-\epsilon_-|<1$, and evaluating the derivative ${\partial \over \partial x}$ acting on $\epsilon_+|x+{z\over 2}|$, and using a similar estimate as before for the funciton inside  of the $x$-integral,
\begin{equation}
  |(3^a)|\leq C'_{10} {\hbar^2 \over \mu} \int dz \Big({\sqrt{2m}\over \hbar}\Big)^2\nu \Big|{\partial \nu \over \partial z}\Big|({\sqrt{2m}\over \hbar} \nu) \int dx e^{-{\sqrt{2m}\over\hbar}\nu |x|} |x|\leq C_{10} {m\over \mu} \nu_0^2
.\end{equation}

Let us now look at the rest,
which contains integrals like 
\begin{equation}
  \int dx \epsilon_\pm \epsilon_\pm |x\pm {z\over 2}|^2
,\end{equation}
by similar arguments they are less than combinations 
\begin{equation}
  \int dx e^{-{\sqrt{2m}\over \hbar} }[ |x|^2+2|x||z|+|z|^2]
,\end{equation}
the leading term of which comes from $|x|^2$ integration.
Thus, to the leading order, we have an upper bound, 
\begin{equation}
|(3^b )| \leq C_{11}' {\hbar^2\over \mu}\int dz \psi^2 \Big({\sqrt{2m}\over \hbar}\nu\Big)\Big({\sqrt{2m}\over \hbar} \Big)^2\Big|{\partial \nu\over \partial z}\Big|^2 \Big({\sqrt{2m}\over \hbar}\nu \Big)^{-3}
\leq C_{11} {m\over \mu} \nu_0^2
.\end{equation}

Estimation of each one of these terms are therefore shown to be smaller than the leading term within this approximation,  as claimed.
\section{Appendix-II}

Here, we show that the short distance behavior of the effective potential for the relativistic particle indeed goes as  $-{1\over 2}|z|$ as claimed in the main text. To do this we need to expand the original term as follows,
\begin{equation}
\int _0^\infty dt e^{t\mu_0}K_0(m\sqrt{t^2+z^2})=\frac1{2}\sum_{k=0}^\infty\frac{\mu_0^k}{k!}\int_0^\infty dt t^k\int_0^\infty \frac{du}{u}e^{-u(t^2+z^2)-\frac{m^2}{4u}}
.\end{equation}
Using the fact that for any finite value of $t$ the exponential is unformly convergent and there is Gaussian suppression for very large values of $t$, we get,
\begin{eqnarray}
\int _0^\infty dt e^{t\mu_0}K_0(m\sqrt{t^2+z^2})&=&\frac1{2}\sum_{k=0}^\infty \frac{\mu_0^k}{k!}\int_0^\infty dt t^ke^{-ut^2}\int_0^\infty \frac{du}{u}e^{-uz^2-\frac{m^2}{4u}}\nonumber\\
&=&\frac1{2}\sum_{k=0}^\infty \frac{\mu_0^k}{k!}\int_0^\infty dt t^ke^{-t^2}\int_0^\infty \frac{du}{u^{\frac{k+1}{2}+1}}e^{-uz^2-\frac{m^2}{4u}}\nonumber\\
&=&\frac1{4}\sum_{k=0}^\infty \Gamma\left(\frac{k+1}{2}\right)\frac{\mu_0^k}{k!}|z|^{k+1}\int_0^\infty \frac{du}{u^{\frac{k+1}{2}+1}}e^{-u-\frac{m^2z^2}{4u}}\nonumber
\end{eqnarray}
We then recognize that the last integrals correspond to  modified Bessel functions, we get,
$$
\int _0^\infty dt e^{t\mu_0}K_0(m\sqrt{t^2+z^2})
=\frac1{2}\sum_{k=0}^\infty \Gamma\left(\frac{k+1}{2}\right)\frac{\mu_0^k}{k!}2^\frac{k+1}{2}\left(\frac{|z|}{m}\right)^\frac{k+1}{2}K_{\frac{k+1}{2}}(m|z|)
.$$
We recall the duplication formula for the gamma function,
\begin{equation}
\Gamma(x)\Gamma\left(x+\frac1{2}\right)=2^{1-2x}\sqrt{\pi}\Gamma(2z)
,\end{equation}
Using this identity the integral expression becomes,
$$
(*)=\int _0^\infty dt e^{t\mu_0}K_0(m\sqrt{t^2+z^2})=\sqrt{\pi}\sum_{k=0}^\infty \frac{\mu_0^k}{2^\frac{k+1}{2}\Gamma\left(\frac{k}{2}+1\right)}\left(\frac{|z|}{m}\right)^\frac{k+1}{2}K_{\frac{k+1}{2}}(m|z|)
$$
To identify the proper $z$ behaviour we need to write the sum on the right side over integer and half-integer terms separately,
$$
(*)=\sqrt{\pi}\sum_{n=0}^\infty \frac{\mu_0^{2n}}{2^{n+\frac1{2}}\Gamma(n+1)}\left(\frac{|z|}{m}\right)^{n+\frac1{2}}K_{n+\frac1{2}}(m|z|)+\sqrt{\pi}\sum_{n=1}^\infty \frac{\mu_0^{2n-1}}{2^n\Gamma\left(n+\frac1{2}\right)}\left(\frac{|z|}{m}\right)^nK_n(m|z|).
$$
We now remind the reader 
the series expantion of modified Bessel functions with integer and half-integer order (see \cite{gradsh}) respectively, 
\begin{eqnarray}
K_{n+\frac1{2}}(x)&=&\sqrt{\frac{\pi}{2x}}e^{-x}\sum_{k=0}^n \frac{\Gamma(n+k+1)}{k!\Gamma(n-k+1)(2x)^k},\nonumber\\
K_n(x)&=&\frac1{2}\sum_{k=0}^{n-1}(-1)^k\frac{(n-k-1)!}{k!\left(\frac{x}{2}\right)^{n-2k}}\nonumber\\
&+&(-1)^{n+1}\sum_{k=0}^\infty \frac{\left(\frac{x}{2}\right)^{n+2k}}{k!(n+k)!}\left[\ln\left(\frac{x}{2}\right)-\frac1{2}\psi(k+1)-\frac1{2}\psi(k+n+1)\right]\nonumber
.\end{eqnarray}
When we use these expansions in the sum, the first part of the sum takes the form, 
\begin{eqnarray}
\Sigma_1&=&\sqrt{\pi}\sum_{n=0}^\infty \frac{\mu_0^{2n}}{2^{n+\frac1{2}}\Gamma(n+1)}\left(\frac{|z|}{m}\right)^{n+\frac1{2}}K_{n+\frac1{2}}(m|z|)\nonumber\\
&=&\pi\sum_{n=0}^\infty\sum_{k=0}^n\frac{\mu_0^{2n}}{2^{n+k+1}k!}\frac{\Gamma(n+k+1)}{\Gamma(n+1)\Gamma(n-k+1)}\frac{|z|^{n-k}}{m^{n+k+1}}e^{-m|z|}\nonumber\\
&=&\pi\sum_{n=0}^\infty\sum_{k=0}^n\frac{\mu_0^{2n}}{2^{n+k+1}k!}\frac{\Gamma(n+k+1)}{\Gamma(n+1)\Gamma(n-k+1)}\frac{|z|^{n-k}}{m^{n+k+1}}(1-m|z|+\frac1{2}m^2z^2+...).\nonumber
\end{eqnarray}
The constant term and $|z|$ term can be found by setting $k=n$ and $k=n-1$, and then identifying each contribution. 
For $k=n$ the first term gives,
$$
\pi\sum_{n=0}^\infty \frac{\mu_0^{2n}}{2^{2n+1}(\Gamma(n+1))^2}\frac{\Gamma(2n+1)}{m^{2n+1}}(1-m|z|+\frac1{2}m^2z^2+...)
,$$
which contains a constant term as well as a $|z|$ term, which we write as,
$$
C_1=\pi\sum_{n=0}^\infty \frac{\mu_0^{2n}}{2^{2n+1}(\Gamma(n+1))^2}\frac{\Gamma(2n+1)}{m^{2n+1}}
$$
$$
C_2|z|=-\pi\sum_{n=0}^\infty \left(\frac{\mu_0}{m}\right)^{2n}\frac{\Gamma(2n+1)}{2^{2n+1}(\Gamma(n+1))^2}|z|
.$$
If we set $k=n-1$,  in the expression for $\Sigma_1$ above, we find,
$$
=\pi\sum_{n=1}^\infty \left(\frac{\mu_0}{m}\right)^{2n}\frac{\Gamma(2n)}{2^{2n}\Gamma(n)\Gamma(n+1)}|z|
.$$
Note that {\it here the sum should begin} from $n=1$.
In these two expressions we use now the duplication formula, and this leads to the cancelations except the $n=0$ term in the first expression.
As a result, the leading orders  of the first summation $\Sigma_1$ becomes, 
$$
\pi\sum_{n=0}^\infty\frac{\mu_0^{2n}}{2^{2n+1}(\Gamma(n+1))^2}\frac{\Gamma(2n+1)}{m^{2n+1}}-\frac{\pi}{2}|z|
.$$
Let us now focus on the second summation, $\Sigma_2$,
$$
\Sigma_2=\sqrt{\pi}\sum_{n=1}^\infty \frac{\mu_0^{2n-1}}{2^n\Gamma\left(n+\frac1{2}\right)}\left(\frac{|z|}{m}\right)^nK_n(m|z|)=\sqrt{\pi}\frac1{2}\sum_{n=1}^\infty\sum_{k=0}^{n-1}(-1)^k\frac{\mu_0^{2n-1}}{2^{2k}\Gamma\left(n+\frac1{2}\right)}\frac{(n-k-1)!}{k!}\frac{|z|^{2k}}{m^{2n-2k}}
$$
$$
+\sqrt{\pi}\sum_{n=1}^\infty\sum_{k=0}^\infty(-1)^{n+1}\frac{\mu_0^{2n-1}}{2^{2k+2n}\Gamma\left(n+\frac1{2}\right)}\frac{m^{2k}|z|^{2n+2k}}{k!(k+n)!}\left[\ln\left(\frac{m|z|}{2}\right)-\frac1{2}\psi(k+1)-\frac1{2}\psi(k+n+1)\right]
.$$
The second expression is of higher order than  $O(z)$ so we can neglect it, thus  we consider only the first expression. The first term also contributes higher order terms except $k=0$  for each $n$, thus we replace it, to order $z$, as follows,
$$
\sqrt{\pi}\frac1{2}\sum_{n=1}^\infty\sum_{k=0}^{n-1}(-1)^k\frac{\mu_0^{2n-1}}{2^{2k}\Gamma\left(n+\frac1{2}\right)}\frac{(n-k-1)!}{k!}\frac{|z|^{2k}}{m^{2n-2k}}\rightarrow\sqrt{\pi} \frac1{2}\sum_{n=1}^\infty \frac{\mu_0^{2n-1}}{\Gamma\left(n+\frac1{2}\right)}\frac{\Gamma(n)}{m^{2n}}
$$
As a result the  total constant contribution becomes, 
$$
\pi\sum_{n=0}^\infty \frac{\mu_0^{2n}}{2^{2n+1}(\Gamma(n+1))^2}\frac{\Gamma(2n+1)}{m^{2n+1}}+\frac{\sqrt{\pi}}{2}\sum_{n=1}^\infty \frac{\mu_0^{2n-1}}{\Gamma\left(n+\frac1{2}\right)}\frac{\Gamma(n)}{m^{2n}}
$$
$$
=\frac{\sqrt{\pi}}{2}\sum_{n=0}^\infty\frac{\mu_0^{2n}}{m^{2n+1}}\frac{\Gamma\left(n+\frac1{2}\right)}{\Gamma(n+1)}+\frac{\sqrt{\pi}}{2}\sum_{n=0}^\infty \frac{\mu_0^{2n+1}}{m^{2n+2}}\frac{\Gamma(n+1)}{\Gamma\left(n+\frac{3}{2}\right)}
$$
$$
=\frac1{m}\frac{\sqrt{\pi}}{2}\sum_{n=0}^\infty\left(\frac{\mu_0}{m}\right)^{2n}\left(\frac{\Gamma\left(n+\frac1{2}\right)}{\Gamma(n+1)}+\frac{\mu_0}{m}\frac{\Gamma(n+1)}{\Gamma\left(n+\frac{3}{2}\right)}\right)=\frac{\arccos\left(-\frac{\mu_0}{m}\right)}{\sqrt{m^2-\mu_0^2}}
,$$
as can be verified by expanding the integral term for $z=0$ (or by setting $z=0$ and identifying this integral as before).
So, as claimed, we find for the integral expression,
\begin{equation}
{1\over \pi}\int_0^\infty dt e^{t\mu_0}[K_0(m\sqrt{t^2+z^2})-K_0(mt)]=-\frac1{2}|z|+O(z^2)
\end{equation}

\section{Appendix-III}

In this short appendix we will demostrate that indeed the main contribution of the integral given in (\ref{key1}) comes from the region where $tu/M$ terms in comparison to $s$ terms are neglected.
To make sense of this claim we rewrite the integral in terms of sclaed out variables, $s\mapsto s/m^2$ and $t\mapsto t/m$. This leads to
\begin{eqnarray}
&\ &\frac{1}{4\pi m}\int [dR]\chi^{\dagger}(R)\exp\left[-{stu\over 2}\Big({m\over M}\Big){1\over tu \frac{m}{2M}+s}\left({R^2\over m^2}\right)\right]\nonumber\\
&\ &\times\int_0^1 du \int_0^\infty dt t^2 e^{-ut[H_0-(\mu_0+\delta' E)]/m}\int_0^\infty ds \frac{e^{-s-\frac{t^2}{4s}}}{s^{\frac3{2}}\sqrt{tu\frac{m}{2M}+s}}\chi(R).
\end{eqnarray}
Note that the term in question is important when $u$ is not too small, if $u<<1$ we can neglect these terms. However for the time being let us not put a restriction on $u$.  If $s$ is large the term in question is negligible, unless $t$ is very large, but then there are exponential $t$ terms to suppress the integrals. Then the  main contributions  may come when $t$  and $s$ are both small.
Note that $t$ should be much smaller in this case since otherwise exponential term $e^{-t^2/4s}$ leads to a high suppression. If $s$ is small, we need at most $t\sim s^{1/2}$. So the terms we want to drop off  may not be negligible if  we have $s^{1/2} m/M\sim s$, that means $s\sim (m/M)^2$.
Let us look at the integral within this interval, note that for $0<t<s^{1/2}$ and $0<s<(m/M)^2$,
the second exponential term  is of order one plus lower order corrections in $m/M$. So we replace it with $1$ to estimate. The first one can be written as,
\begin{equation}
\exp\left[-{stu\over 2}\Big({m\over M}\Big){1\over tu \frac{m}{2M}+s}\left({R^2\over m^2}\right)\right]=\exp\left[-s{tu\over s}\Big({m\over 2M}\Big){1\over \frac{tu}{s} \frac{m}{2M}+1}\left({R^2\over m^2}\right)\right],
\end{equation}
which is of the form $\exp [-\alpha\frac{x}{1+x}]$ and this negative exponential is maximized when the function of $x$ is minimized, hence we set $x=0$.
That means the exponential could be replaced by $1$. Hence the integral has an upper bound in the interval of interest,
\begin{eqnarray}
\frac{1}{4\pi m}\int [dR]\chi^{\dagger}(R)\chi(R)
\int_0^1 du \int_0^{(m/M)^2} ds\int_0^{s^{1/2}}  t^2 dt \frac{e^{-s-\frac{t^2}{4s}}}{s^{\frac3{2}}\sqrt{tu\frac{m}{2M}+s}}.\nonumber
\end{eqnarray}
By replacing the upper limit of the $t$ integral with $m/M$ and using Cauchy-Schwartz inequality, replacing $e^{-s}$ by $1$,  we estimate that the integral expression without the orthofermion part (which gives a projection) is smaller than,
 \begin{eqnarray}
\int_0^1 du \int_0^{m/M}  t^2 dt\left[ \int_0^{(m/M)^2} ds\frac{e^{-\frac{t^2}{2s}}}{s^{3}}\right]^{1/2}\left[\int_0^{(m/M)^2} {ds\over{tu\frac{m}{2M}+s}}\right]^{1/2}.\nonumber
\end{eqnarray}
Now we estimate as follows,
\begin{eqnarray}
&\ & \int_0^1 du \int_0^{m/M}  t^2 dt\left[ \int_0^{(m/M)^2} ds\frac{e^{-\frac{t^2}{2s}}}{s^{3}}\right]^{1/2}\left[\int_0^{(m/M)^2} {ds\over{tu\frac{m}{2M}+s}}\right]^{1/2}\nonumber\\
&\ & =\int_0^1 du \int_0^{m/M}  t^2 dt\left[ {1\over t^4}\int_0^{(m/M)^2/t^2} ds\frac{e^{-\frac{1}{2s}}}{s^{3}}\right]^{1/2}\left[\ln\Big({tu\frac{m}{2M}+(m/M)^2\over tu\frac{m}{2M}}\Big)\right]^{1/2}\nonumber\\
&\ &\leq \int_0^1 du \int_0^{m/M} dt \left[ \int_0^{\infty} ds\frac{e^{-\frac{1}{2s}}}{s^{3}}\right]^{1/2}\left[\ln\Big({tu\frac{M}{2m}+1\over tu\frac{M}{2m} }\Big)\right]^{1/2}\nonumber\\
&\ & ={m\over M}\left[ \int_0^{\infty} ds\frac{e^{-\frac{1}{2s}}}{s^{3}}\right]^{1/2}\int_0^1 du \int_0^{1} dv \ln^{1/2}\Big({{uv/2}+1\over uv/2}\Big)\nonumber\\
&\ &\leq C{m\over M},
\end{eqnarray}
hence unimportant at this level as claimed.

\section{Acknowledgment}
O. T. Turgut would like to express his deep gratitude to  Jens Hoppe for discussions and the kind invitation to KTH, Stockholm, where parts of this work are completed. O. T. Turgut also would like to thank F. Erman and L Akant for discussions.

\end{document}